\documentclass{ar-1col-S2O}
\usepackage[numbers]{natbib}
\usepackage{url}
\usepackage{amsmath}
\usepackage{amssymb}
\newcommand{\beq}{\begin{equation}}
\newcommand{\eeq}{\end{equation}}
\newcommand{\bea}{\begin{eqnarray}}
\newcommand{\eea}{\end{eqnarray}}

\newcommand{\dd}{\mathrm{d}}

\newcommand{\pT}{p_{\rm T}}
\newcommand{\jpsi}{J/\psi}
\newcommand{\psip}{\psi(2{\rm S})}
\newcommand{\chic}{\chi_c}
\newcommand{\YiS}{\ensuremath{\Upsilon(1\mathrm{S})}}
\newcommand{\YiiS}{\ensuremath{\Upsilon(2\mathrm{S})}}
\newcommand{\YiiiS}{\ensuremath{\Upsilon(3\mathrm{S})}}
\newcommand{\YS}{\ensuremath{\Upsilon}}
\newcommand{\raa}{R_{\mathrm{AA}}}

\newcommand{\sqrtsNN}{\sqrt{s_{\mathrm{NN}}}}
\newcommand{\qqbar}{\ensuremath{\mathrm{q\bar{q}}}}
\newcommand{\ccbar}{\ensuremath{\mathrm{c\bar{c}}}}
\newcommand{\bbbar}{\ensuremath{\mathrm{b\bar{b}}}}
\newcommand{\dsigmacc}{\ensuremath{\mathrm{d}\sigma_{\mathrm{c\bar{c}}}/\mathrm{d}y}}
\newcommand{\meanNp}{\ensuremath{\langle N_{part}\rangle}}

\newcommand{\vtwo}{v_{2}}

\jname{Xxxx. Xxx. Xxx. Xxx.}
\jvol{AA}
\jyear{YYYY}
\doi{10.1146/((article doi))}

\begin{document}

% Page header
\markboth{Andronic, Arnaldi}{Quarkonia and Deconfined Matter}

% Title
%\title{Title: Subtitle}
\title{Quarkonia and Deconfined Quark-Gluon Matter in Heavy-Ion Collisions}

%Authors, affiliations address.
\author{Anton Andronic$^1$ and Roberta Arnaldi$^2$
\affil{$^1$Institute for Nuclear Physics, University of M\"unster, Germany, 48159; email: andronic@uni-muenster.de}
\affil{$^2$INFN Sezione di Torino, Turin, Italy; email: arnaldi@to.infn.it}}

%Abstract
\begin{abstract}
In this report, we present an experimental overview of quarkonium results obtained in nucleus-nucleus collisions, with a focus on the data collected at the LHC. We discuss the current understanding of charmonium and bottomonium behavior in the deconfined medium produced in such collisions, comparing the various observables now accessible to state-of-the-art theoretical models. We also discuss the open points and how future heavy-ion experiments aim to clarify these aspects.
\end{abstract}

%Keywords, etc.
\maketitle

%Table of Contents
\tableofcontents

% Heading 1
\section{INTRODUCTION}

The suppression of quarkonium production was proposed almost forty years ago~\cite{Matsui:1986dk} as a unique and unambiguous signature of the formation of a plasma of quarks and gluons (QGP) in collisions of heavy nuclei at high energies \cite{Busza:2018rrf,Harris:2024aov}. 
The original idea by T.\,Matsui and H.\,Satz consisted in the assumption that the color screening (the so-called ``Debye screening''), induced by the presence of a hot and dense medium, prevents the binding of the $\rm q$ and $\bar {\rm q}$ quarks forming the quarkonium, leading to a significant reduction of its production. 
This screening becomes stronger and stronger as the temperature of the system increases, hence the suppression is predicted to be sequential, first affecting the more loosely bound quarkonium states and only when higher temperatures are reached, also the most tightly bound states~\cite{Digal:2001ue,Karsch:2005nk}. 
\begin{marginnote}[]
\entry{QGP}{quark-gluon plasma is a very hot and dense medium, predicted by QCD, where quarks and gluons are no more confined into hadrons}
\end{marginnote}

The theory paper of Matsui and Satz immediately triggered the experimental search for this signature. Starting from 1986, the first studies were carried out at CERN SPS, with the NA38 experiment, investigating S-U collisions, followed shortly after by the NA50 experiment, exploiting Pb-Pb interactions, and later on, at the beginning of the year 2000, by NA60 with In-In collisions. The charmonium resonances $\jpsi$ and $\psip$ (i.e. the $\rm{c}$ and $\bar{\rm c}$ quarks S-wave pair) were the first quarkonium states measured at the top SPS energy, at the center-of-mass energy per nucleon-nucleon $\sqrtsNN$ = 17.3 GeV. Charmonium was observed to be ``anomalously suppressed'', i.e. suppressed well beyond the modification induced by cold nuclear matter effects~\cite{NA50:2004sgj}, leading, together with other observations carried on at the SPS, to the CERN announcement, in the year 2000, of the evidence of QGP formation in Pb-Pb collisions~\cite{CERNpressrelease}. The $\psip$ yields in AA collisions were also observed to be more reduced with respect to the $\jpsi$ ones, suggesting the expected sequential order in the suppression~\cite{NA50:2006yzz}.

\begin{marginnote}[]
\entry{Debye screening}{the phenomenon where the QGP medium screens the quark-antiquark binding potential. The screening length, known as Debye length, is inversely proportional to the QGP temperature.}
\end{marginnote}

The CERN SPS results were then followed, at the beginning of the new century, by corresponding measurements at higher center-of-mass energies, firstly at RHIC at BNL (Au--Au collisions at top $\sqrtsNN$ = 200 GeV) and later on, in 2010, at the CERN LHC (Pb--Pb collisions at top $\sqrtsNN$ = 2.76 TeV in the so-called LHC Run1, reaching 5.02 TeV in Run2 and most recently 5.36 TeV in Run3). These higher available energies allowed to extend further the study of the QGP properties.
It was soon discovered, for example, that, with the increase of the collision energy, new mechanisms play a role.  The observed suppression of the quarkonium yields, compared to the production in pp collisions, turns out to be (partially) balanced by (re)generation mechanisms~\cite{BraunMunzinger:2000csl,Thews:2000rj} related to the large amount of $\ccbar$ quark pairs present in a deconfined medium (the number of $\ccbar$ pairs, $N_{\ccbar}$, produced on average in a central collision is about 0.2 at the top SPS energy, $\sim$10 at RHIC and reaches $\sim$120 at the top LHC energy\,\cite{BraunMunzinger:2000csl,Andronic:2006ky}). This feature was predicted to occur at the colliders in 2000~\cite{BraunMunzinger:2000csl,Thews:2000rj}. Quarks and antiquarks close in phase space can in fact combine to form a quarkonium state, either during the deconfined phase and/or at the hadronization of the system, leading to an increase of the charmonium yields.

\begin{marginnote}[]
\entry{(Re)generation}{mechanism leading to the production of quarkonia due to the combination of ${\rm q}$ and ${\rm \bar q}$ (from the same or a different initially-produced pair) close in phase space. 
}
\end{marginnote}

The increase in collision energy brought further benefits to quarkonium studies. In fact, one of the peculiarities of quarkonia is that these particles can come in a large variety of states, characterized by different binding energies and, therefore possibly differently affected by the medium created in the heavy-ion collisions. While SPS results were limited to the $\jpsi$ and $\psip$, the increase in $\sqrtsNN$ up to the LHC energies opened up the study of the heavier bottomonia, i.e. bound states of $\rm{b}$ and $\bar{\rm b}$ quarks. These measurements, previously limited by the low production cross section, now nicely complement the charmonium results. 

Nowadays, at RHIC and LHC all the existing experiments pursue quarkonium measurements over (slightly) different rapidity ($\rm y$) and transverse momentum ($\pT$) regions. Hence, measurements over a broad kinematic coverage are now available, allowing very detailed quarkonium studies and putting severe constraints on the theory models. 

The availability of experimental results over a broad kinematic range and a wide range of collision energies, coupled with the possibility of studying a full family of particles, certainly put quarkonium in an ideal position to investigate the deconfined QGP medium, confirming its pivotal role, since now already forty years, as a signature of QGP formation.

\section{THEORETICAL\ BACKGROUND}

\subsection{The vacuum properties of quarkonia}

We list in \textbf{Table~\ref{tab:prop}} the basic vacuum properties of the quarkonium states, both \ccbar\ and \bbbar\ families, which are currently employed for the study of the QGP (for all, $J^{PC}=1^{--}$). Here, $\Delta m$ is the difference between the mass of the $\mathrm{D},\bar{\mathrm{D}}$ or $\mathrm{B},\bar{\mathrm{B}}$ pairs, namely the lightest mesons which contain the $\mathrm{c}$,$\mathrm{\bar{c}}$ or $\mathrm{b}$,$\mathrm{\bar{b}}$ quarks, respectively, and the mass of the respective quarkonium state. This quantity is a proxy for the binding energy of a given quarkonium state.

\begin{table}[htb]
  \centering
  \caption{The vacuum properties of selected quarkonium states, the branching ratios (B.R.) into lepton pairs, and the relative yield from feed-down (F.D.) from higher-mass quarkonium states.}\label{tab:prop}
\begin{tabular}{|l|c|c|c|c|c|} \hline
State & Mass (GeV) & $\Delta m$ (GeV) & B.R. & F.D. \\ \hline \hline
J/$\psi$ & 3.097 & 0.633 & 5.97\% & 30\% \\ \hline
$\psip$ & 3.686 & 0.044 & 0.8\% & n.a. \\ \hline \hline
\YiS & 9.46 & 1.1 & 2.4\% & 24\%  \\ \hline
\YiiS &  10.02 & 0.54 & 1.9\% & 24\%  \\ \hline
\YiiiS &  10.36 & 0.21 & 2.2\% & 40\%  \\ \hline \hline
\end{tabular}
\end{table}

Also listed are the branching ratios (B.R.) for the dielectron or dimuon decays, the channels in which the quarkonia are usually (in heavy-ion collisions exclusively) detected and the average fraction originating from feed-down (F.D.) through decays from higher-mass states, either $\mathrm{S}$ or $\mathrm{P}$ states, in pp collisions \cite{Lansberg:2019adr,Boyd:2023ybk}. 
For instance, in case of J/$\psi$ about 8\% originates from F.D. from $\psip$ and about 22\% from the $\chi_{c1,2}$ $\mathrm{P}$ states \cite{Lansberg:2019adr}.
We note that the F.D. fractions are $p_\mathrm{T}$-dependent \cite{Lansberg:2019adr,Boyd:2023ybk}, listed in \textbf{Table~\ref{tab:prop}} are approximate average values. We note that, for AA collisions, where the relative production rates of the various quarkonium species (within a family) are measured to be different than in pp collisions, these F.D. components will be different as well.
For $\jpsi$ and $\psip$ mesons, an additional F.D. contribution of 10-15\% and 20-30\%, respectively, originates on average from hadrons containing bottom quarks; this contribution has a strong dependence on transverse momentum, being more important at high $\pT$~\cite{ALICE:2021edd,LHCb:2021pyk}.

\subsection{General considerations}

Inclusive heavy-quark pair (\qqbar, where $\mathrm{q}$ denotes either charm of bottom) production is a perturbative QCD (hard/partonic) process, occurring in the initial phase of a nucleus-nucleus collision. 
The timescale of the process is of the order of $1/2m_{\mathrm{q}}$, with $m_\mathrm{q}$ the mass of the charm and bottom quarks, about 1.3 and 5 GeV/$c^2$, respectively. Consequently, for both charm and bottom quarks this timescale is well below 1\,fm/$c$. Quarkonium, a bound \qqbar\ state, is produced over a timescale of 1\,fm/$c$ or larger since it involves the separation of the \qqbar\ pair to build the wave function of a respective quarkonium state. 
In heavy-ion collisions, the quarkonium formation time is comparable to (at the collider energies larger than) the time needed for the thermalization of the quark-gluon matter and the development of the collective expansion. The charm and bottom quarks are expected to thermalize more slowly than the bulk of the quark-gluon medium, formed by the light quarks and the gluons. 

The heavy-quark potential in QGP is made quantitative through Lattice QCD (LQCD) calculations, where the complete information is encoded in the quarkonium spectral functions, see Refs.~\cite{Mocsy:2013syh,He:2022ywp} for reviews.
The spectral functions contain information on the quarkonium binding energies and the inelastic reaction rates as a function of the temperature of the medium. The melting temperatures for various quarkonium states can also be derived, albeit in a model/procedure-dependent way.
Recent LQCD calculations show that the real part of the potential, which embodies the classical screening picture, exhibits essentially no temperature dependence, while the imaginary part, which implies dissociative (collisional) processes, strongly depends on temperature~\cite{Bazavov:2023dci}.
Given these considerations, the simple and attractive picture of the ``quarkonium thermometer'' for QGP~\cite{Matsui:1986dk,Mocsy:2013syh,Kluberg:2009wc}, which is based on the screening of the real part of the potential, appears currently too simplistic.
Instead, quarkonia are studied to infer their in-medium properties and the facets of the strong interaction in hot/dense QCD matter that lead to the dissociation and (re)generation processes~\cite{Rothkopf:2019ipj}.
For instance, the long-range confining force was shown to be clearly modified in the QCD medium, based on the theoretical description of bottomonium data in a transport model employing a potential model\,\cite{Du:2019tjf}.

\begin{marginnote}[]
\entry{Rapidity}{ $y=\frac{1}{2}\ln\frac{E-p_z}{E+p_z}$, where $E$ is the energy and $p_z$ the longitudinal momentum of a particle. Quantifies the distribution of produced particles in the longitudinal (beam) direction.}
\end{marginnote}

The treatment of quarkonium production in theoretical approaches is intrinsically related to that of the open heavy-flavor  hadrons.
Of the inclusive heavy-quark production yields only a small fraction will hadronize as quarkonia.
For instance, at the LHC, in a central (0-10\% class) Pb-Pb collision about 120 \ccbar\ pairs are produced on average in initial collisions in the full phase space. At midrapidity ($y=0$) there are about 16 \ccbar\ pairs measured on average in one unit of rapidity, of which about 0.7\% are measured in $\jpsi$ mesons~\cite{ALICE:2022wpn}.
Heavy quarks are overall essential probes of the QGP \cite{He:2022ywp,Dong:2019byy,Apolinario:2022vzg,Das:2024vac}. 
The diffusion process of heavy quarks in the QGP plays a crucial role in theoretical modeling of quarkonium production and interaction in the hot/dense QGP\cite{He:2022ywp}.

\subsection{Theoretical models}
\label{sec:theoretical_models}
Previous reviews on quarkonium production in heavy-ion collisions \cite{Mocsy:2013syh,Rothkopf:2019ipj,Zhao:2020jqu,Rapp:2008tf} focused intensively on the theoretical aspects. As our review is more focused on the experimental status, we briefly review here only the main lines of current theoretical effort on the description of quarkonium in QGP. A recent comprehensive inter-comparison of various models is available in Ref.~\cite{Andronic:2024oxz}.
There are two basic approaches for the theoretical description of quarkonium in QGP: i) the statistical hadronization model and ii) dynamical models, based on transport approaches using either semiclassical kinetic-rate equations or an open quantum systems framework.

\subsubsection{The Statistical Hadronization Model}
In the Statistical Hadronization Model for charm (SHMc) \cite{BraunMunzinger:2000csl,Andronic:2006ky,BraunMunzinger:2009dzl,Andronic:2021erx} full dissociation of all quarkonium states in the QGP and exclusive generation at the QCD (crossover) phase boundary is assumed.
This implies for heavy quarks a hadronization process which is concurrent with that of lighter quarks and gluons. At high collision energies hadronization coincides with the chemical freeze-out stage in the evolution of the hot QCD medium and with the QCD chiral crossover transition \cite{Andronic:2017pug}, which LQCD predicts at a temperature $T_{pc}=156-158$\,MeV \cite{HotQCD:2018pds,Borsanyi:2020fev}.

The complete thermalization of the heavy quarks in the expanding deconfined medium down to QCD crossover transition is an essential condition for the applicability of SHMc.
In this approach, the knowledge of the inclusive \qqbar\ production cross-section along with the chemical freeze-out (hadronization) temperature $T_{cf}\simeq 157\pm 2$\,MeV obtained from the analysis of the yields of hadrons composed of light valence quarks \cite{Andronic:2017pug}, is sufficient to determine the total ($\pT$-integrated) yield of all hadrons containing heavy quarks in ultra-relativistic nuclear collisions.
\begin{marginnote}[]
\entry{QCD chiral crossover transition}{is the transition between hadronic matter and QGP. According to LQCD, at baryochemical potential $\mu_B\sim0$ the transition is a crossover at $T_{pc}=156-158$\,MeV.}
\end{marginnote}

\begin{marginnote}[]
\entry{Chemical freeze-out}{is the stage, in the evolution of heavy-ion collisions, where the relative abundances of the different particle species are fixed.}
\end{marginnote}

This is based on the balance equation relating the initial inclusive charm production to the yields of hadronic states:
\begin{equation}
    N_{\ccbar} = \frac{1}{2} g_c V \sum_{i} n^{{\rm th}}_i \frac{I_1(N_{c}^{tot})} {I_0(N_{c}^{tot})} \,
    + \, g_c^2 V \sum_{j} n^{{\rm th}}_j \, + \, \frac{1}{2} g_c^2 V \sum_{k} n^{{\rm th}}_k \frac{I_2(N_{c}^{tot})} {I_0(N_{c}^{tot})},
  \label{eq:balance}
\end{equation}
where $N_{\ccbar}\equiv \dd N_{\ccbar}/\dd y$ denotes the rapidity density of charm quark pairs produced in initial hard collisions and the (grand-canonical) thermal densities for open and hidden charm hadrons are given by $n_{i,j,k}^{{\rm th}}$. The index $i$ runs over all open charm states with one valence charm or anti-charm quark ($\mathrm{D}, \mathrm{D}_s, \Lambda_c, \Xi_c, \Omega_c$ and antiparticles), the index $j$ over all quarkonium states ($\jpsi, \chi_c, \psip$), and the index $k$ over open charm states with two charm or anti-charm quarks ($\Xi_{cc}, \Omega_{cc}$ and antiparticles). 
The fugacity factor $g_c$ is obtained by solving \textbf{Equation~\ref{eq:balance}} for a given collision centrality class and enters in the model predictions of the yields of hadrons with charm quarks and antiquarks linearly for single-charm hadrons and quadratically for charmonia and doubly-charmed baryons.
The ratio of the modified Bessel functions, $I_\alpha/I_0$, is a (canonical) correction for the exact conservation of charm~\cite{Cleymans:1990mn,Gorenstein:2000ck}. The argument,  
$N_{c}^{tot}$, is the total charm content (particles and antiparticles), consequently containing, besides the thermal densities of charmed hadrons and the volume, the $g_c$ factor.
The thermal densities are computed in the grand canonical ensemble using the latest version of the SHMc~\cite{Andronic:2017pug,Andronic:2019wva}, with the chemical freeze-out temperature $T_{cf}\simeq 157$ MeV. The fireball volume per unit rapidity at mid-rapidity is $V = 4997 \pm 455$\,{fm}$^3$  for the most central 10\% Pb-Pb collisions at LHC energy $\sqrtsNN$ = 5.02 TeV. In this case, based on the measured average value $N_{\ccbar} \simeq 16$ for one unit of rapidity~\cite{ALICE:2022wpn}, $g_c\simeq 31.5$. While the thermal densities do not vary with centrality of the collision, $N_{\ccbar}$ and $V$ in \textbf{Equation~\ref{eq:balance}} are centrality-dependent and scale with the number of nucleon-nucleon collisions, $N_{coll}$, and number of participating nucleons, $N_{part}$, respectively\footnote{Centrality is estimated based on data and the Glauber model~\cite{dEnterria:2020dwq}, a geometric nuclear overlap model, and expressed either as a range in the total geometric cross section or as the average number of participating nucleons, \meanNp, for a given range.}. This leads to a quasi-linear dependence of $g_c$ on $N_{part}$.
Thermal charm production as well as charm quark-antiquark annihilation in Pb-Pb collisions are neglected, as they were estimated to be very small at the LHC energies and negligible for lower energies~\cite{Andronic:2006ky,Braun-Munzinger:2000uqj}.

With the assumption of the kinetic freeze-out taking place also at the QCD phase boundary and employing hydrodynamics, the transverse momentum distributions can be calculated as well~\cite{Andronic:2023tui}. A corona contribution from the dilute periphery of the fireball is added, both for the total and the $\pT$-differential yields, based on measurements in pp collisions.
The SHM was applied to the bottom sector too~\cite{Andronic:2006ky,Andronic:2022ucg}, although in this case incomplete thermalization of bottom quarks needs to be considered.

\subsubsection{Transport Models}
\begin{figure}[hbt]
\begin{tabular}{lr} \begin{minipage}{.49\textwidth}
\includegraphics[width=.95\linewidth]{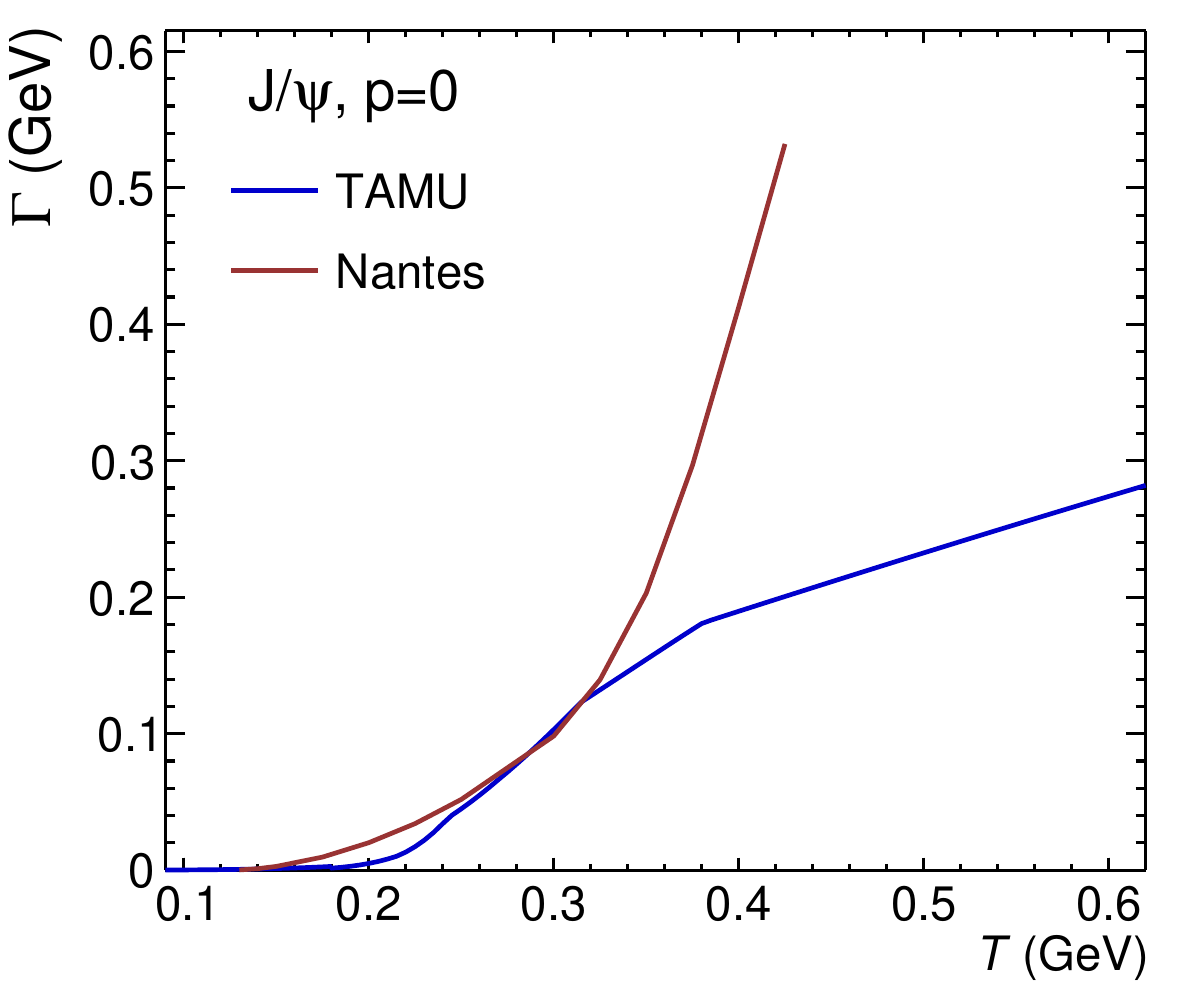}
\end{minipage} &\begin{minipage}{.49\textwidth}
\includegraphics[width=.95\linewidth]{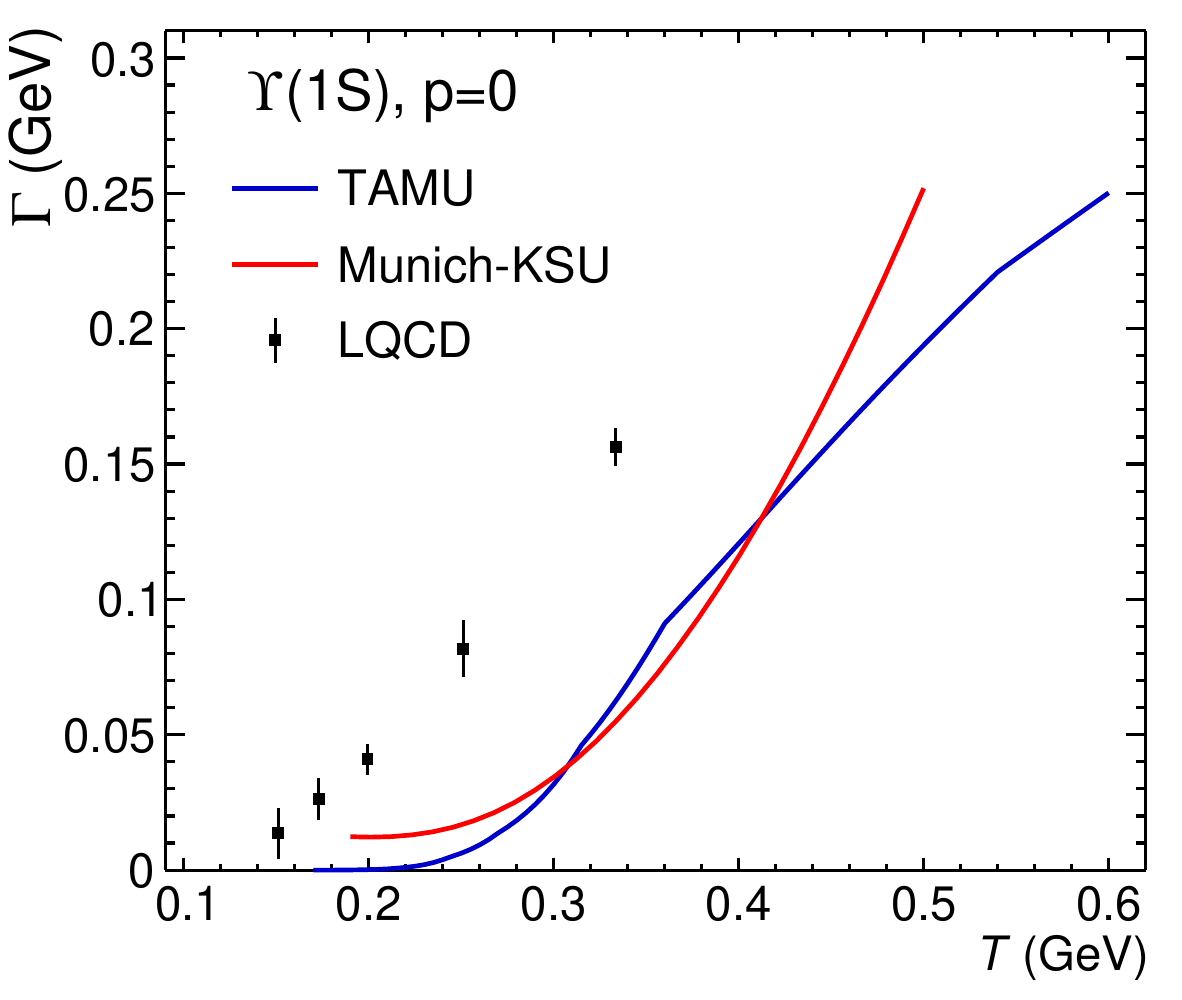}
\end{minipage} \end{tabular}
\caption{The temperature dependence of the reaction rates $\Gamma$ for the $\jpsi$ (left panel) and $\YiS$ mesons (right panel, note the different scale of the vertical axis) at zero momentum. The values used in the semi-classical (TAMU~\cite{Wu:2024gil}) and open-quantum systems (Nantes\,\cite{Delorme:2024rdo} and Munich-KSU\,\cite{Brambilla:2023hkw}) transport approaches are shown, compared, in case of $\YiS$, with LQCD predictions~\cite{Larsen:2019zqv}.}
\label{fig:gamma}
\end{figure}

In the semi-classical transport \cite{Thews:2000rj,Grandchamp:2001pf,Wu:2024gil}, the time evolution of the yield $N$ for a given quarkonium species (charmonium and bottomonium) is governed by a rate equation:
\begin{equation}
\frac{\dd N(\tau(T))}{\dd\tau} = -\Gamma(T(\tau))\left[N(\tau)-N^{eq}(T(\tau))\right], \label{eq:rate}
\end{equation}
where $\tau$ is the time in the expanding fireball reference frame, 
$T(\tau)$ is the local temperature, $\Gamma$ is the reaction rate and $N^{eq}$ is the equilibrium value for the yield. The loss term in this equation, $-\Gamma(T(\tau))N(\tau)$, describes the suppression of initially-produced quarkonia, while the other term is a gain term and corresponds to the (re)generation of quarkonia from \qqbar\ pairs in the medium. Here, $\mathrm{q}$ and $\mathrm{\bar{q}}$ can originate from the same initial \qqbar\ pair (a so-called ``diagonal term'') or from two different initial pairs (``non-diagonal term'').
The dependence $T(\tau)$ is fixed through hydrodynamical model comparison to collective
flow data at a given collision energy and for a given centrality class. For instance, for central Pb--Pb collisions (0-10\%) at $\sqrtsNN=5.02$\, TeV, $T$ decreases from about 600 MeV at the very early equilibrium stage ($\tau\simeq0.5$\, fm/$c$) to $T_{pc}$ at the QCD crossover boundary in about 15\,fm/$c$~\cite{Andronic:2024oxz}.
The reaction rate $\Gamma$ needs to be modeled phenomenologically. The state-of-the-art  modeling employs in-medium ($T$-dependent) quark masses and binding energies\,\cite{Tang:2024dkz}, matched to LQCD results~\cite{Larsen:2019zqv}. For instance, in the thermodynamic T-matrix approach~\cite{Wu:2024gil,Tang:2024dkz}, the in-medium $\mathrm{c}$-quark mass decreases by about 20\% up to $T=400$ MeV and the binding energies of quarkonia decrease strongly as a function of $T$. 
Other LQCD-anchored approaches\,\cite{Brambilla:2024tqg} employ no medium-modified quark masses or binding energies, see Ref.\,\cite{Andronic:2024oxz} for a comprehensive overview.
The reaction rates increase significantly with $T$ (see \textbf{Figure\,\ref{fig:gamma}}) and show, in general, an increase with momentum. 
Note that \textbf{Equation~\ref{eq:rate}} is written in the ``natural units" system, in which $\hbar=c=1$. In this system of units $\Gamma$ is GeV, and the conversion to S.I. units is: 1\,GeV $\simeq$ 1.5$\times$10$^{24}$\,s$^{-1}$.

\vspace{0.5cm}

It was observed in the study of Ref.~\cite{Andronic:2024oxz} that the model inputs for the reaction rates, as well as for the in-medium binding energy of the various quarkonium states, are quite different among the various model implementations, a situation that clearly needs to be improved upon. 
We illustrate in \textbf{Figure\,\ref{fig:gamma}} the temperature dependence for the reaction rates for the ground-states of charmonia and bottomonia.
The reaction rates are larger for the excited states, where the spread between the values used in various models is also larger.
All implementations include (re)generation, including the non-diagonal term. 

Besides such transport models based on rate equations and/or semiclassical Boltzmann equations~\cite{Rapp:2017chc}, transport approaches utilizing open-quantum system (OQS) frameworks have been developed in the recent years, see the review in Ref.~\cite{Akamatsu:2020ypb}. 
The quantum master equations can be reduced, under certain conditions, mainly in case the binding energy of the quarkonium state is (much) smaller than the temperature of the medium, to the so-called Lindblad equation. This is called the regime of the quantum Brownian motion, see discussion in Ref.~\cite{Delorme:2024rdo}.
The hierarchy of times discussed above (and, equivalently, of energy scales) plays a role in this procedure and leads to domains of applicability of the OQS approaches~\cite{Brambilla:2024tqg,Delorme:2024rdo}. The hierarchy is also exploited in an effective theory of the strong interaction, the (potential) non-relativistic Quantum Chromodynamics, (p)NRQCD~\cite{Brambilla:1999xf}, which allows the phenomenological application of OQS to quarkonium production or suppression in the QGP~\cite{Brambilla:2024tqg,Akamatsu:2020ypb}.
In current quantum-transport approaches, which mostly focus on bottomonia, (re)generation is within a single \bbbar\ pair (diagonal term) and shown to be significant~\cite{Brambilla:2023hkw}. 
The quantum effects appear to be significant mostly for the early stages of quarkonium evolution~\cite{Andronic:2024oxz}. 

\begin{figure}[htb]
\begin{tabular}{lr} \begin{minipage}{.49\textwidth}
\includegraphics[width=1.\linewidth]{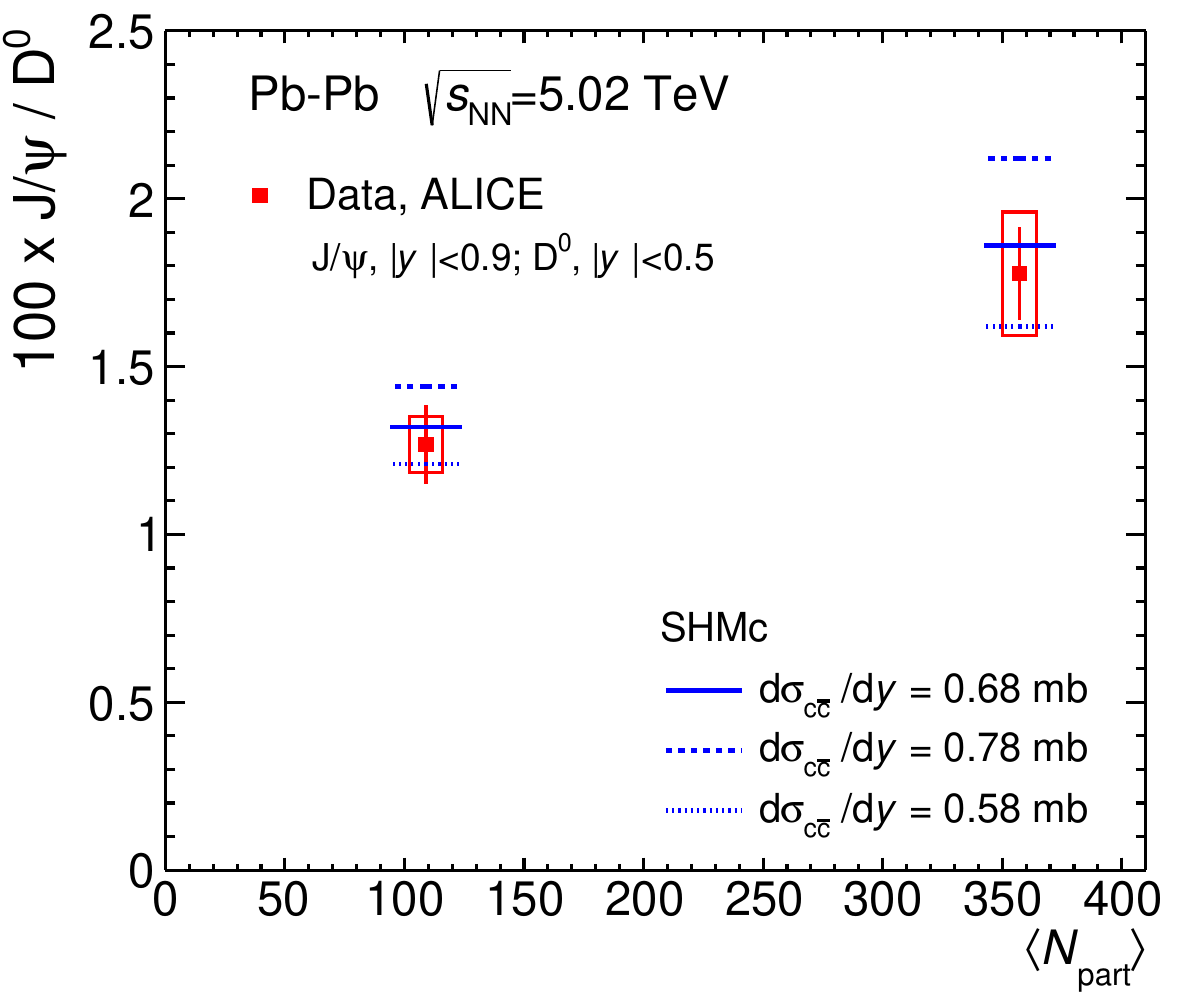}
\end{minipage} &\begin{minipage}{.49\textwidth}
\includegraphics[width=1.\linewidth]{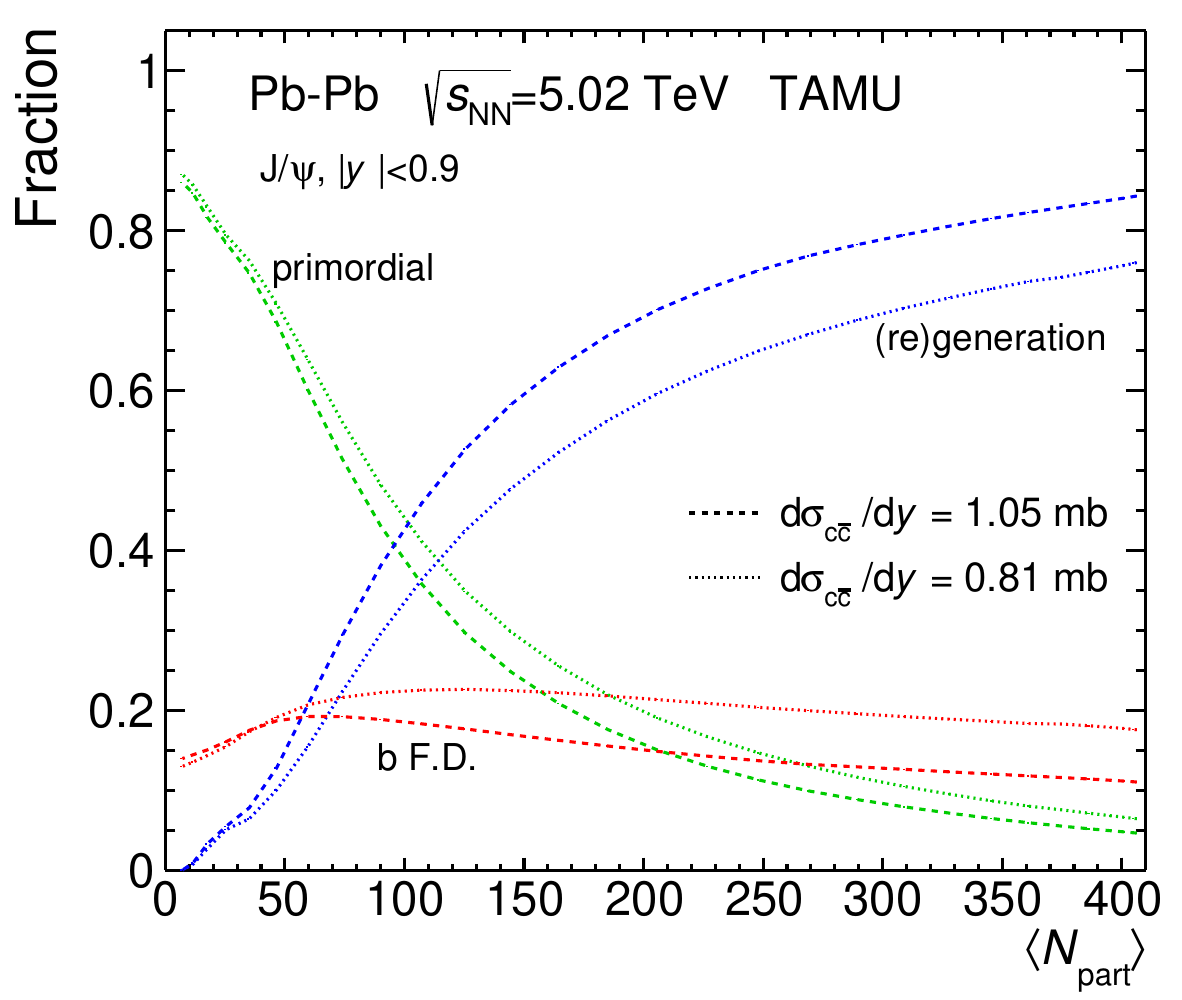}
\end{minipage} \end{tabular}
\caption{Left panel: the yield ratio $\jpsi$ to D$^0$ (in percent) as a function of centrality. The data from ALICE \cite{ALICE:2023gco,ALICE:2021rxa} are compared to the prediction of SHMc \cite{Andronic:2021erx}. Right panel: the fraction of primordial, (re)generation and F.D. from bottom hadrons for $\jpsi$ yields as a function of centrality, calculated in the TAMU transport model~\cite{Wu:2024gil}.}
\label{fig:jpsiD0}
\end{figure}

In \textbf{Figure\,\ref{fig:jpsiD0}} we illustrate the features of the SHMc and TAMU transport model concerning $\jpsi$ production in Pb-Pb collisions at $\sqrtsNN$ = 5.02 TeV. 
The SHMc describes very well the relative production of the $\jpsi$ to $\mathrm{D}^0$ mesons (left panel). 
The uncertainty in the model, arising from the charm yield  $\dd N_{\ccbar}/\dd y$ and fixed through the experimentally measured yields of $\mathrm{D}^0$ mesons in 0-10\% Pb-Pb collisions\,\cite{ALICE:2021rxa}, is largely correlated with that of the experimental data. A remarkable feature of the SHMc is that it describes the absolute yield of the $\jpsi$ mesons, alongside that all the measured open-charm hadrons~\cite{ALICE:2022wpn}.
The TAMU transport model predicts (right panel of \textbf{Figure\,\ref{fig:jpsiD0}}) that the fraction of $\jpsi$ mesons produced through (re)generation has a centrality dependence and reaches about 80\% of the total $\jpsi$ yield in central collisions, where the initially-produced $\jpsi$ mesons contribute only around 5\%. 
The situation is different in mid-central collisions, where for $\meanNp \simeq$ 100 the TAMU model predicts a fraction of $\jpsi$ yield from (re)generation in QGP equal to the primordial yield, while for the SHMc all yield is newly generated at the QCD crossover boundary. Here, the contrast between the two models is plain.
For both models, a larger inclusive charm production cross section \dsigmacc\ leads to a more pronounced (re)generation of $\jpsi$ mesons, seen for the TAMU model directly and for the SHMc meson as a larger ratio to the $\mathrm{D}^0$ yield.
An early prediction of the transport models was that the (re)generation component is predominantly at low $\pT$\,\cite{Zhao:2011cv} and it leads to collective flow\,\cite{Zhou:2014kka}.

\begin{marginnote}[]
\entry{Nuclear shadowing}{the reduction, or saturation, of parton densities (sistribution functions) in nuclei compared to free nucleons~\cite{Klasen:2023uqj}.}
\end{marginnote}
Not only for the SHMc and TAMU models illustrated here, but in general \dsigmacc\ is a fundamental input for the theoretical calculations for charmonium (and to some extent also bottomonium) production in QGP.
The values of \dsigmacc\ considered in the calculations shown in \textbf{Figure\,\ref{fig:jpsiD0}} are in fact spanning the uncertainties used in the two models for \dsigmacc.
Note that this is the equivalent value for nucleon-nucleon collisions, after nuclear shadowing for Pb-Pb collisions was factored in. 
The SHMc and TAMU models employ two different values, \dsigmacc = 0.68\,$\pm$\,0.10 mb and \dsigmacc = 0.93\,$\pm$\,0.12 mb, respectively. 
The SHMc value is derived from the experimentally measured yields of $\mathrm{D}^0$ mesons in 0-10\% Pb-Pb collisions\,\cite{ALICE:2021rxa} and assuming hadronization fractions as in SHMc, with an enhanced spectrum of charged baryons\,\cite{Andronic:2021erx} (see also\,\cite{He:2019tik,ALICE:2023sgl}). The value in the TAMU model is derived from the measured value in pp collisions and assuming shadowing in the range 10-30\%. This current mismatch in the input values of the two models needs to be resolved.
A precise experimental evaluation of this quantity requires the measurement of all ground-state open charm mesons and baryons down to zero $\pT$ in nuclear collisions and remains a challenge to date.

Below, we compare the predictions of the SHMc, the TAMU semiclassical transport
model, and the Munich-KSU OQS+pNRQCD model with current experimental data, with
a focus on the measurements at the LHC.
Several other groups have made contributions and achieved a good description of (some of) the quarkonium data in the semiclassical transport\,\cite{Zhou:2014kka,Ferreiro:2012rq,Ferreiro:2018wbd,Song:2023zma,Wolschin:2020kwt}
or quantum approaches\,\cite{Blaizot:2021xqa,Yao:2020xzw,Villar:2022sbv,Chen:2024iil}.
We note that not all these models are based on the LQCD constraints mentioned above~\cite{Andronic:2024oxz}.
A comprehensive comparison of CMS data on $\Upsilon$ production at the LHC with theoretical models is available in Ref.\,\cite{CMS:2023lfu}.

\section{EXPERIMENTAL\ OVERVIEW AND COMPARISON TO THEORY} 
As discussed, quarkonium has been one of the first observables investigated in heavy-ion collisions and the understanding of its behavior in hot and very dense matter has gone hand in hand with the evolution of the experiments at the more and more powerful accelerators. In this report, we will outline the status of quarkonium measurements in nucleus-nucleus collisions, with emphasis on the RHIC and LHC data. A review of the SPS and early RHIC data and their theoretical interpretation is also available in Ref.\,\cite{Kluberg:2009wc,BraunMunzinger:2009dzl}. 

\subsection{Quarkonium observables}
Several observables are used to study the quarkonium behavior in heavy-ion interactions.
The most common one is the so-called nuclear modification factor, $\raa$. In the $\raa$, defined as $\raa = N_{\jpsi}^{AA} / (N_{\jpsi}^{pp} \times {\langle N_{coll} \rangle})$, the quarkonium production yield in AA collisions, $N_{\jpsi}^{AA}$, is compared to the quarkonium yield in proton-proton interactions, $N_{\jpsi}^{pp}$, scaled by the average number of collisions $\langle N_{coll} \rangle$. 

With the availability of data from RHIC and LHC, other observables, often requiring large statistics to be studied, became of interest for quarkonium studies. As an example, observables such as the elliptic flow $\vtwo$ and the polarization, provide additional information on the quarkonium behavior in heavy-ion collisions and, in particular, on the interplay between suppression and (re)generation mechanisms.

Details on the dynamics of the early stages of the collisions can, for example, be inferred from the azimuthal particle distribution. In fact, in non-central collisions, the geometrical overlap region, and hence the initial matter distribution is anisotropic. If the matter is strongly interacting, this spatial anisotropy converts, through multiple collisions, into an anisotropic momentum distribution of the emitted particles. The beam axis and the impact parameter vector of the colliding nuclei define the reaction plane and the second coefficient ($\vtwo$) of the Fourier expansion of the particle azimuthal distribution with respect to this plane is called the elliptic flow. Charm quarks, if thermalized in the QGP, will flow together with the lighter particles, leading to a quarkonium $\vtwo$ different from zero. 

Finally, quarkonia may also exhibit polarization, defined as the alignment of the particle spin with respect to a chosen axis~\cite{Faccioli:2010kd}. The degree of polarization can be influenced by the presence of the QGP, because of the existence of a strong magnetic field in the early stage of its formation and/or because of large vorticity in the QGP. Also, in this case, $\jpsi$ produced by (re)generation might be affected differently than the primordial ones.

\subsection{pp and pA collision systems for quarkonium studies}
\label{sec:pppA}
In heavy-ion experiments, quarkonium is usually studied not only in AA collisions (mainly Pb--Pb collisions at both SPS and LHC and Au--Au at RHIC), but also in pp and pA (at SPS and LHC) or dA interactions (at RHIC) interactions. These collision systems are not extensively covered here, but are reviewed in Ref.~\cite{Andronic:2015wma}. 

Quarkonium measurements in pp are essential to the investigation of its hadronic production mechanism. Furthermore, since in pp collisions no QGP is expected to be formed, pp measurements represent also a  reference for AA results, as can be deduced from the $\raa$ definition. The determination of the pp baseline has to be obtained at the same energy and in the same kinematic range as the AA data. 

Quarkonium measurements in pA or dA collisions are essential to investigate the so-called cold nuclear matter effects (CNM)\footnote{Cold Nuclear Matter effects are initial or final state effects that modify the quarkonium yield. Their presence is related to the nuclear medium.  CNM effects are always present in AA collisions, but they are usually investigated in minimum bias pA collisions, since in this lighter system hot matter QGP effects are considered to be negligible.}. In fact, already from the first quarkonium studies at CERN SPS~\cite{NA50:2006rdp,NA50:2003fvu,NA60:2010wey}, it was immediately realized that the quarkonium production yields are not only influenced by the hot matter formation but also by effects associated to the presence of a nuclear medium. 
Nuclear shadowing, i.e. the modification of the quark and gluon structure function for nucleons inside nuclei (see e.g. Ref. ~\cite{Klasen:2023uqj,Eskola:2016oht,Kovarik:2015cma}) or the formation of a Color Glass Condensate (CGC)~\cite{Iancu:2003xm} involving low-x quarks and gluons, can indeed affect the quarkonium production in nuclear collisions. In addition to these purely initial state effects, both the incoming partons and the $\rm{c}\bar{c}$ pair propagating through the nucleus may lose energy by gluon radiation~\cite{Arleo:2012rs} at various stages of the charmonium formation process, inducing further modifications in the observed yields. 
Finally, the fully formed quarkonium could also be dissociated via inelastic interactions with the surrounding nucleons. This final-state process, which has a dominant role among CNM effects at low collision energy~\cite{NA50:2003pvd,PHENIX:2012czk}, is negligible at the LHC, where the crossing time of the two nuclei is much shorter than the formation time of the resonance~\cite{Blaizot:1987ha,Hufner:1989fn,Kharzeev:1999bh}.

The CNM effects introduced above are present in AA collisions, together with the QGP ones, but can be investigated more directly in minimum bias pA (dA) collisions, where the contribution of QGP effects are thought to be negligible. A precise assessment of the size of CNM effects is hence mandatory, to properly establish the impact of the QGP ones, as it will be discussed in Section~\ref{sec:wrapup}. 

Quarkonium in pA has been extensively studied at LHC~\cite{ALICE:2019qie,ATLAS:2017prf,CMS:2022wfi,LHCb:2018psc} and RHIC~\cite{PHENIX:2022nrm,PHENIX:2019brm,STAR:2021zvb}. While the ground states production, the $\jpsi$ and the $\YiS$, is dominated by shadowing or energy loss~\cite{Albacete:2017qng,Vogt:2019jxd,Strickland:2024oat}, the interpretation of the excited states is more complex, with some theoretical models even suggesting the presence of a deconfined medium in high-multiplicity p-Pb collisions at LHC\,\cite{Strickland:2024oat,Du:2018wsj}.
Initial CNM effects, connected to the creation of the heavy-quark pair, are in fact expected to be of a very similar size for the $\jpsi$ and the $\psip$, given that the two resonances are rather close in mass, however final-state effects, as the break-up of the resonance via interactions with the nucleons of the Pb nuclei, could in principle impact differently the two states, as observed at SPS energies. At LHC (or even RHIC) energies, these final-state effects are negligible, because the quarkonium formation time is larger than the time spent by the $\ccbar$ pair in the medium (at LHC energies the latter ranges between $10^{-4}$ and $7\cdot10^{-2}$ fm/$c$ at low $\pT$\,\cite{ALICE:2022wpn,ALICE:2016sdt,ALICE:2014cgk,ALICE:2020vjy}, while estimates for the formation time range between 0.05 and 0.15 fm/$c$~\cite{Arleo:1999af,McGlinchey:2012bp}).

Unexpectedly, the $\psip$ $R_\mathrm{pA}$ measurement shows a different behavior compared  to the $\jpsi$\,\cite{ALICE:2014cgk,ALICE:2020vjy,LHCb:2016vqr,PHENIX:2016vmz}, with a larger suppression, in particular in the backward rapidity region. This observation is suggestive of additional final-state effects, such as interactions in a dense medium (of hadronic or partonic origin\,\cite{Strickland:2024oat,Du:2018wsj,Ma:2017rsu,Ferreiro:2014bia}), meaning that the medium in high multiplicity p(d)-A collisions might impact the yields of the loosely bound $\psip$.

%%%%%%%%%%%%%%%%%%
\subsection{Charmonium experimental results}
%%%%%%%%%%%%%%%%%%%

$\jpsi$ was the first quarkonium state to be extensively investigated in heavy-ion collisions. 
The first, very precise, charmonium results at CERN SPS ($\sqrtsNN = 17.3$ GeV) have shown a clear additional suppression of the $\jpsi$ yields in central Pb--Pb collisions when compared to pure CNM effects~\cite{NA50:2004sgj, Arnaldi:2008zz}. 
Due to the unavailability of pp collisions, the $\jpsi$ yields were studied in comparison to the Drell-Yan process, which, having an electromagnetic nature, is not affected by the QGP formation.
The size of the observed suppression ($\sim$30\% in the most central interactions) was interpreted as quantitatively consistent with the melting of the weakly bound $\psip$ and $\chic$ states whose feed-down contributions to the $\jpsi$ are globally of the same order. A hierarchy between the $\jpsi$ and $\psip$, with the latter experiencing stronger suppression, was indeed observed by NA50~\cite{NA50:2006yzz}\footnote{"Anomalous $\bf{J/\psi}$ suppression" is how the NA50 experiment referred to the first observation of the suppression of the $\jpsi$ yields in Pb-Pb collisions, beyond the modification induced by CNM effects.}.
Nevertheless, on the nature of the medium formed in central Pb--Pb collisions at SPS energies, whether it was a deconfined one or a dense hadron gas, several discussions were triggered
~\cite{BraunMunzinger:2000csl,Gorenstein:2000ck,Grandchamp:2001pf,Kharzeev:1996yx,Capella:2000zp,Blaizot:1996nq}.

With a ten times higher center-of-mass energy ($\sqrtsNN=200$ GeV), the measurements of PHENIX and STAR at RHIC, confirmed the strong suppression of the $\jpsi$ production in Au--Au collisions, compared to pp~\cite{PHENIX:2006gsi,PHENIX:2011img,STAR:2013eve,STAR:2019fge}. The observed suppression has a similar magnitude as the one measured at the SPS and it shows, for the first time, a significant rapidity dependence, being stronger at forward rapidity (1.2 $< |y| <$ 2.2) than at midrapidity ($|y| <$ 0.35).
The interpretation of the RHIC results, assuming both the suppression of the tightly bound $\jpsi$ and a contribution from (re)generation mechanisms, is not fully conclusive. 

The eagerly awaited LHC measurements, with a further increase in $\sqrtsNN$ by a factor initially 14, and nowadays 25, with respect to the RHIC energies and the possibility to cover a very broad rapidity range, down to very low transverse momentum, were indeed ideally suited to clarify the observations made so far. These data were expected to shed light on the role played by (re)generation, as the ultimate test of QGP formation.

\subsubsection{$\jpsi$ in AA collisions}
\begin{figure}[htb]
\begin{tabular}{lr} \begin{minipage}{.49\textwidth}
\includegraphics[width=1.\linewidth]{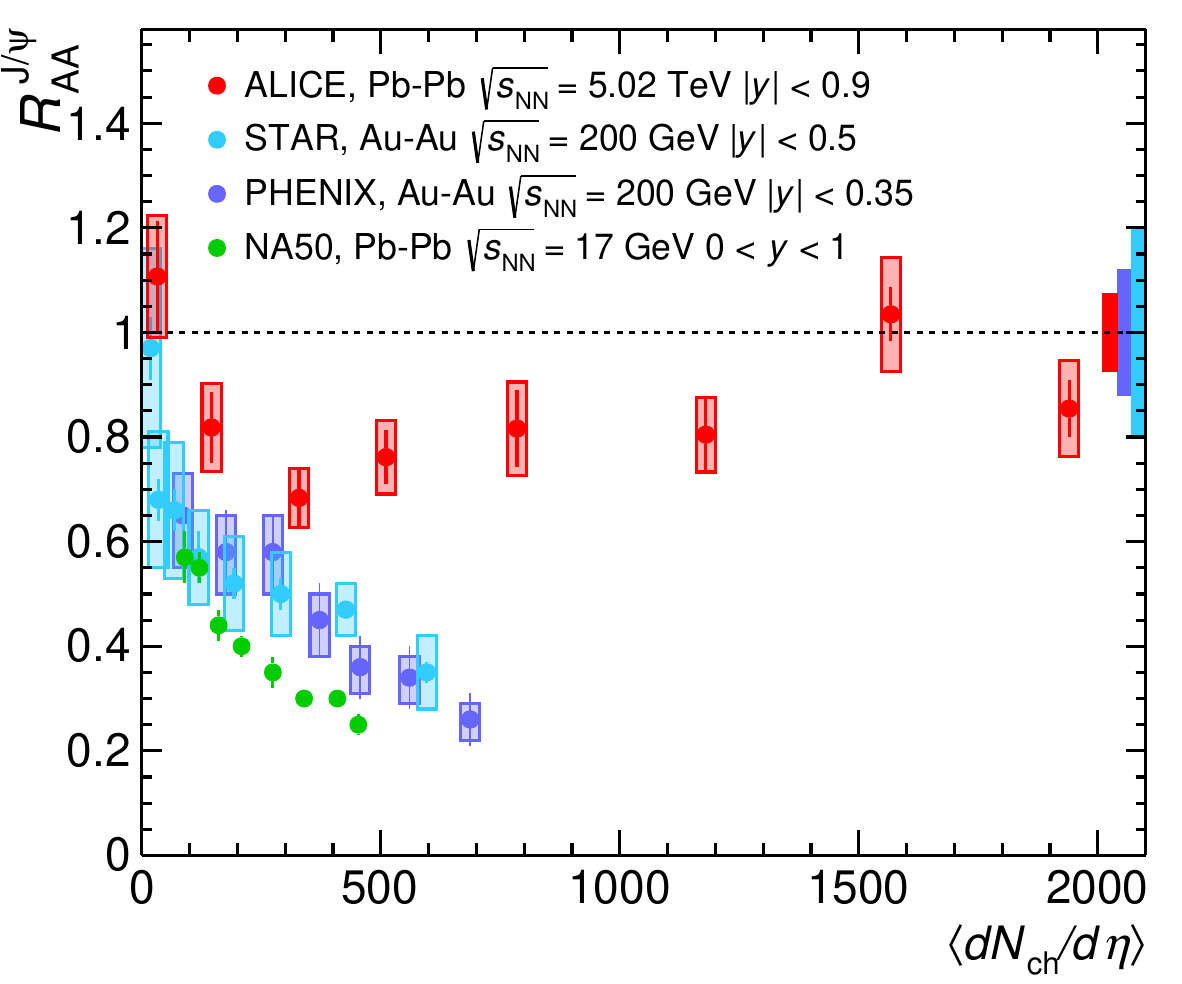}
\end{minipage} &\begin{minipage}{.49\textwidth}
\includegraphics[width=1.\linewidth]{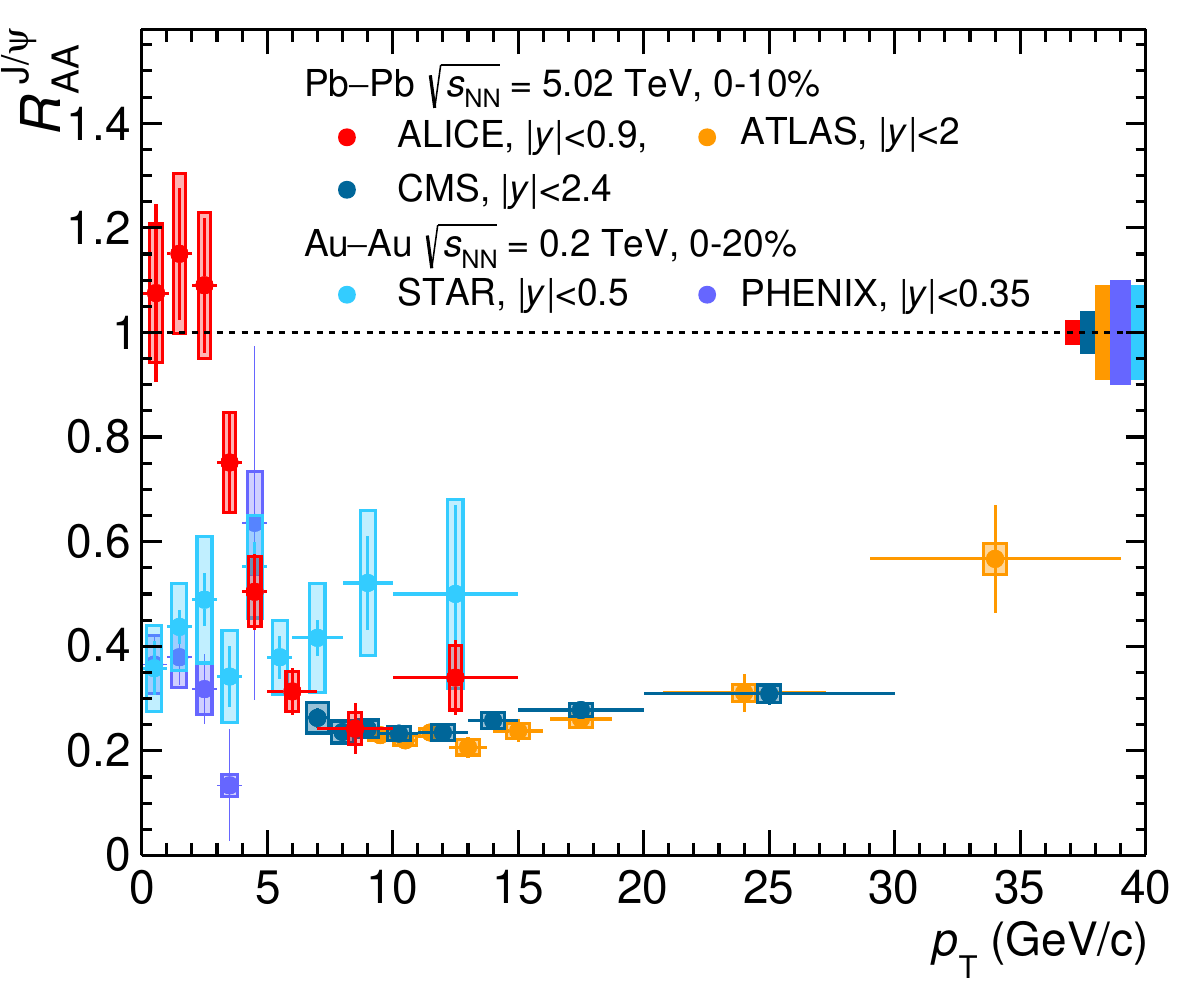}
\end{minipage} \end{tabular}
\caption{$\jpsi$ $\raa$ values for Pb--Pb collisions at $\sqrtsNN$ = 5.02 TeV~\cite{ALICE:2023gco,CMS:2017uuv,ATLAS:2018xms} and for Au--Au collisions at $\sqrtsNN$ = 200 GeV~\cite{PHENIX:2006gsi,STAR:2019fge} is shown as a function of the charged particle multiplicity 
(left panel) and as a function of $\pT$ (right panel). In the left panel, the SPS data from NA50~\cite{NA50:2004sgj} (as shown in Ref. \cite{ALICE:2022wpn}) are also included. Results from NA50, PHENIX, STAR and ALICE refer to inclusive $\jpsi$ production, while those from ATLAS and CMS are for prompt $\jpsi$. However, since the contribution of bottom-hadron feed-down becomes relevant only towards high $\pT$~\cite{ALICE:2023hou}, the comparison between these measurements remains meaningful.} 
\label{fig:exp-jpsi-allsqrts}
\end{figure}

The compilation of the $\jpsi$ results obtained at midrapidity, at all the available nucleon-nucleon center-of-mass energies, from SPS up to LHC, is shown in \textbf{Figure~\ref{fig:exp-jpsi-allsqrts}}. 

In the left panel, results are shown as a function of the mean charged particle multiplicities $\langle$d$N_{\rm {ch}}$/d$\eta$$\rangle$ measured at $\eta$ = 0. In each collision system, this quantity is directly related to the centrality of the interactions and it is a useful variable to compare different collision systems since it is roughly proportional to the initial energy density. At SPS and RHIC energies, the $\jpsi$ $\raa$ decreases when moving from peripheral to central collisions, with a hint of stronger reduction in the SPS results. When moving to the higher LHC energies, two striking features appear in the $\raa$ pattern. On one side the
suppression effects are almost vanishing, in particular towards central collisions, where the $\raa$ reaches unity (it should be pointed out that at LHC central collisions correspond to roughly three times more charged particles than those measured at RHIC). On the other side, while SPS and RHIC results show a strong centrality dependence, the $\jpsi$ $\raa$ measured at LHC shows a very mild dependence, with a slight increase towards the most central collisions, opposite to what was observed at lower energies.
It should be noted that in the ALICE results, a very low $\pT$ cut is applied ($p_{T}>0.3$ GeV/c) in order to exclude the $\jpsi$ from photoproduction processes~\cite{ALICE:2021gpt}, which contribute significantly to the $\jpsi$ yield in particular in peripheral collisions.

In the right panel of \textbf{Figure~\ref{fig:exp-jpsi-allsqrts}}, the $\raa$ evaluated at midrapidity, in the most central events, is shown as a function of the $\jpsi$ $\pT$. Two prominent features can be noticed. First, the LHC  results exhibit a strong $\pT$ dependence. Results from different LHC experiments show an excellent agreement in the common $\pT$ range, with a significant suppression appearing when moving from the low to the high $\pT$ region. 
Second, the RHIC results have a very different trend, with $\raa$ significantly lower than unity, with almost no $\pT$ dependence. The difference with respect to the LHC results is particularly striking in the low $\pT$ region, where the LHC data show even a hint of exceeding unity. It should be noted that this behaviour was not expected in a scenario where color screening was the only mechanism at play.
From both the centrality and the $\pT$ dependence, it is clear that the $\jpsi$ $\raa$ behavior strongly depends on the collision energy. The observation of strong differences between low and high energy results, in particular in central collisions and at low transverse momentum, suggests that different mechanisms set in when the $\sqrtsNN$ increases. 

Since the distribution of charm quarks in the medium is expected to peak around  $y\simeq 0$, reaching its maximum at low $\pT$, the $\jpsi$ produced via (re)generation processes are expected to reflect such distribution, presenting a strong kinematic dependence when studied versus $\pT$ and $y$. 
 Hence, the possibility of measuring charmonium down to zero transverse momentum and in two rapidity regions, a midrapidity ($|y|<0.9$) and a forward one ($2.5< y<4$), puts the ALICE experiment~\cite{ALICE:2023gco,ALICE:2016flj} in an ideal condition to explore the role of (re)generation in Pb--Pb collisions at LHC energies.
 
\begin{figure}[hbt]
\begin{tabular}{lr} \begin{minipage}{.49\textwidth}
\includegraphics[width=1.\linewidth]{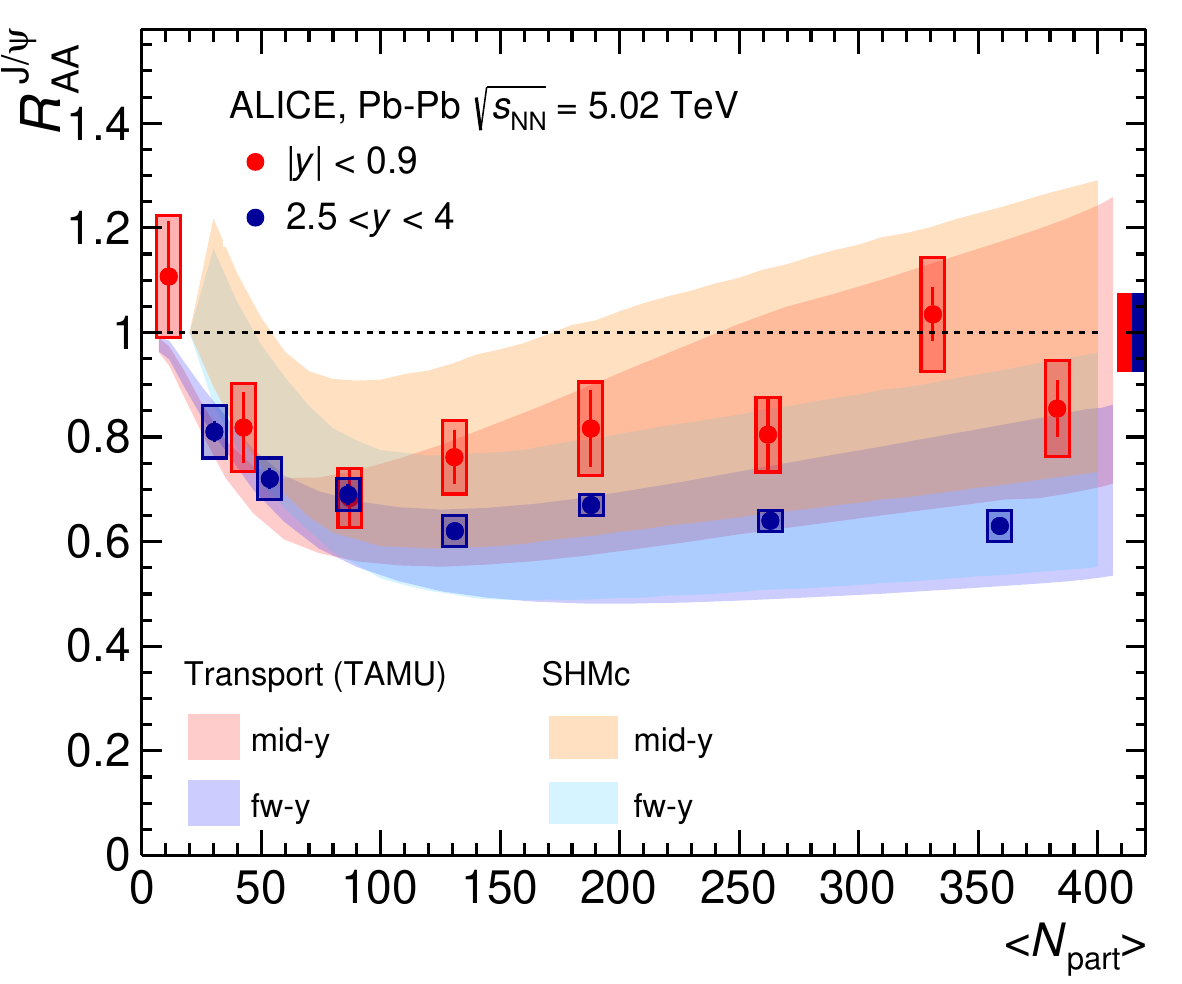}
\end{minipage} &\begin{minipage}{.49\textwidth}
\includegraphics[width=1.\linewidth]{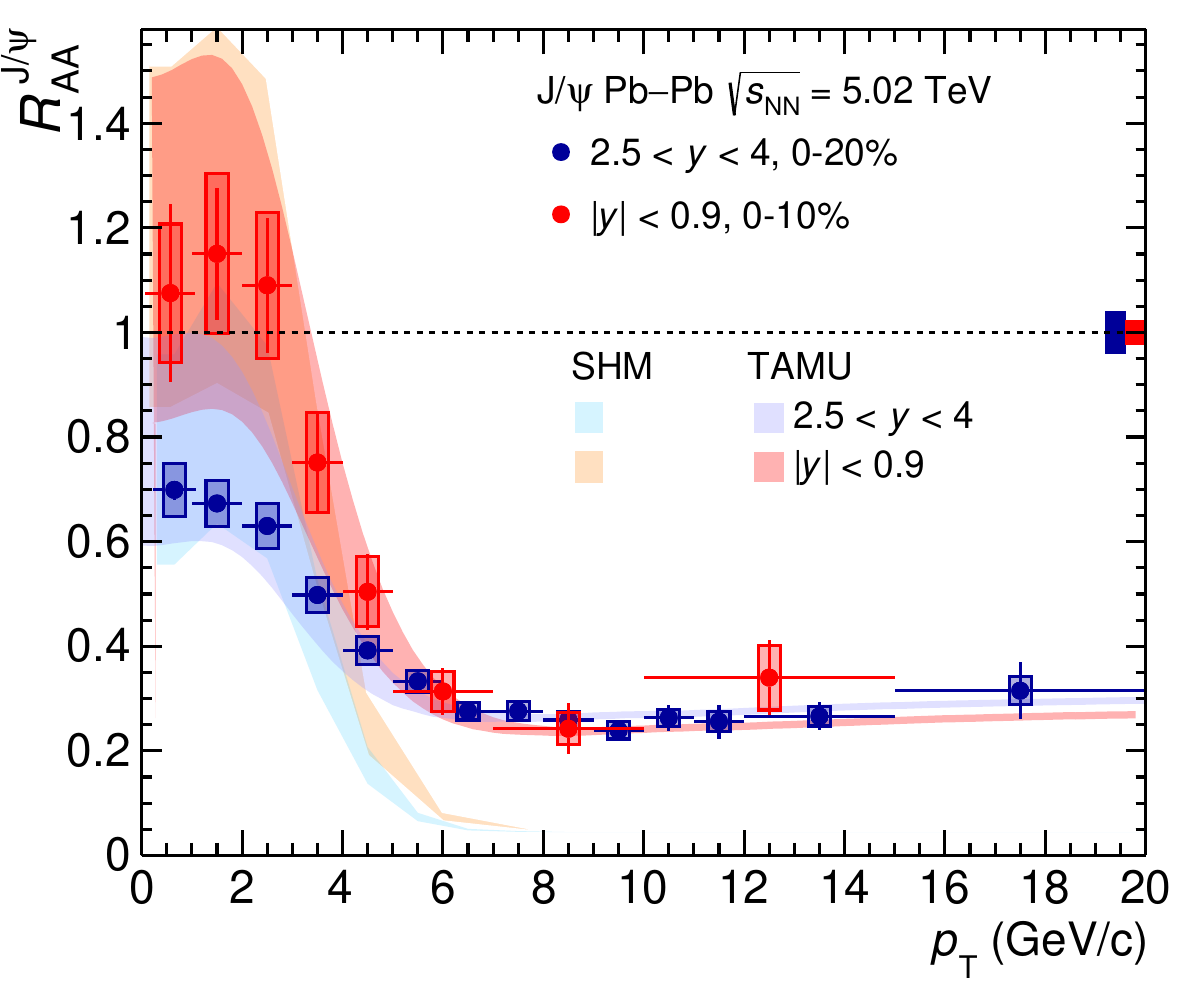}
\end{minipage} \\
\begin{minipage}{.49\textwidth}
\includegraphics[width=1.\linewidth]{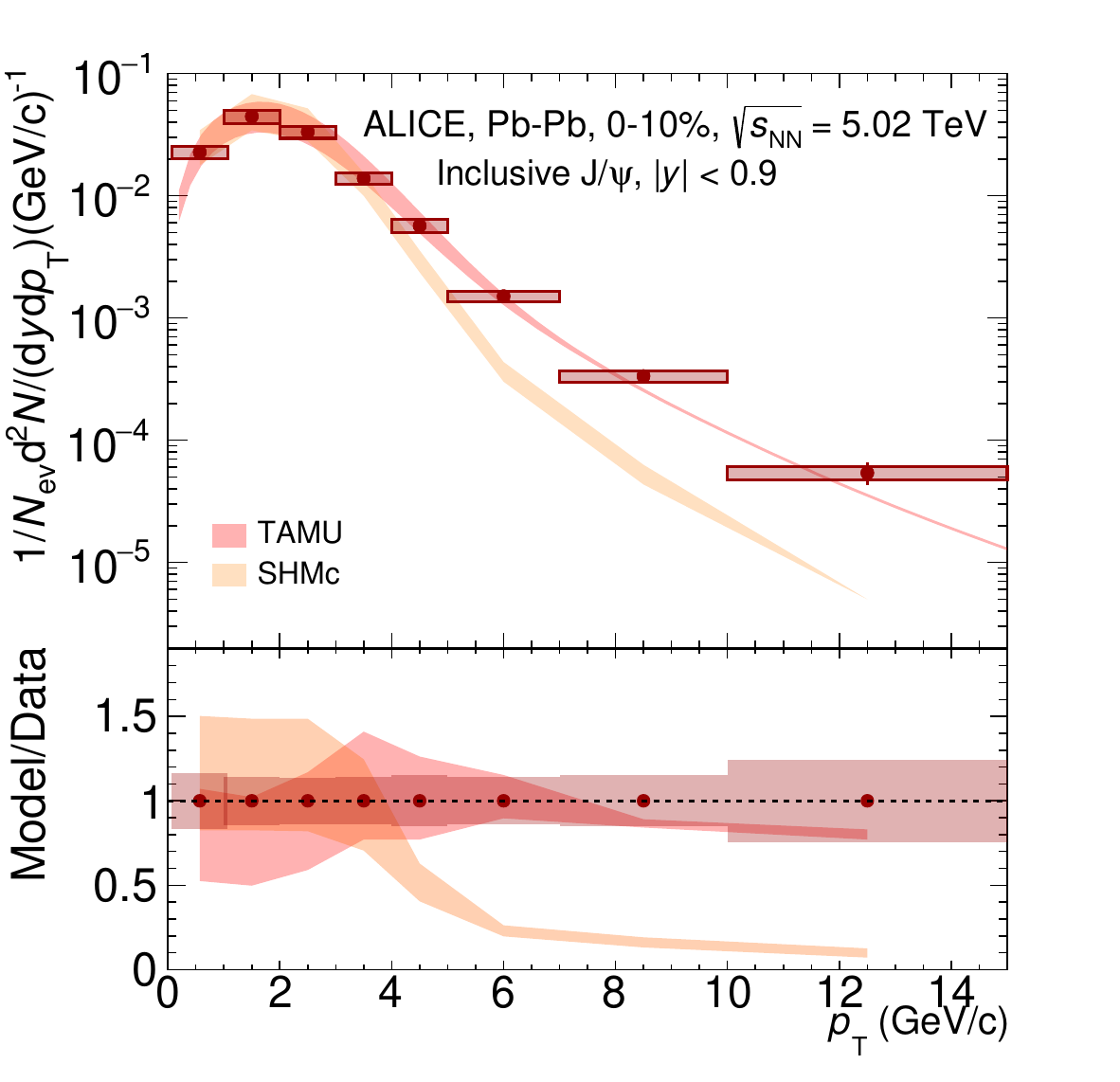}
\end{minipage} &\begin{minipage}{.49\textwidth}
\includegraphics[width=1.\linewidth]{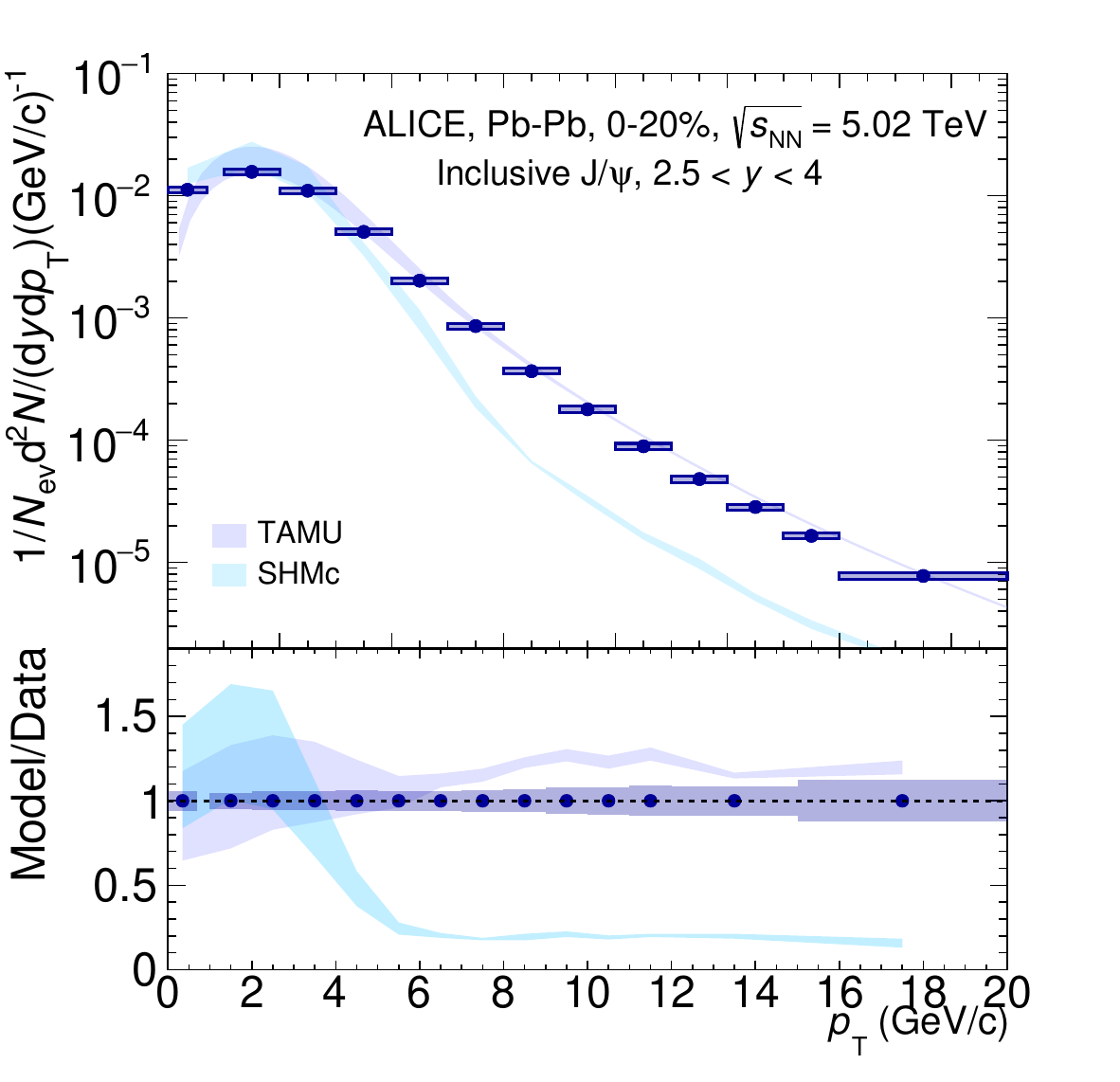}
\end{minipage} \end{tabular}
\caption{The ALICE $\jpsi$ $\raa$ in $|y|<0.9$ and $2.5<y<4$ is shown as afunction of $\langle N_{\rm{part}}\rangle$(top left) and  $\pT$, in the most central collisions (top right). Data are compared to SHMc\,\cite{Andronic:2019wva} and TAMU\,\cite{Wu:2024gil}  calculations. 
In the bottom panels, the fully corrected inclusive $\jpsi$ $\pT$-differential yields are shown, at midrapidity (left) and at forward rapidity (right), in comparison to the same theory calculations. The ratio between data and models is also shown. The filled boxes around unity represent the quadratic sum of statistical and systematic uncertainties from the measurements. }
\label{fig:exp-jpsi-pt_npart}
\end{figure}

The $\jpsi$ $\raa$, measured in both rapidity regions, is shown in \textbf{Figure\,\ref{fig:exp-jpsi-pt_npart}} as a function of centrality ($\langle N_{\rm{part}}\rangle$, top left), and as a function of $\pT$ (top right). 
It shows a strong rapidity dependence, visible when studied both as a function of centrality and $\pT$. More in detail, the differences between the $\raa$ computed in the two rapidity regions become significant in central Pb--Pb collisions and for $\pT <4$\,GeV/$c$, where the $\raa$ measured at midrapidity is close to (or even slightly exceeds) unity. These observations can be interpreted as due to the (re)generation process, dominant, as previously discussed, at low and intermediate $\pT$. 

 A more quantitative assessment of the role of the suppression and (re)generation mechanisms requires a comparison of the measured $\jpsi$ $\raa$ to theoretical models, as also shown in \textbf{Figure\,\ref{fig:exp-jpsi-pt_npart}}.
Both the SHMc\,\cite{Andronic:2019wva} and the TAMU\,\cite{Wu:2024gil} 
calculations nicely describe the centrality dependence of the $\raa$ and the low $\pT$ region, while discrepancies in the high $\pT$ range are visible, in the case of the SHMc model. As discussed in Section~\ref{sec:theoretical_models}, the TAMU approach includes suppression and (re)generation mechanisms in a hot QGP and assumes a 20\% further suppression induced by nuclear shadowing as the dominant cold nuclear matter effect.  The model describes the data, with $\sim$80\% of the $\jpsi$ produced by (re)generation in central collisions at midrapidity, while this number decreases to $\sim$50\% at forward rapidity. At very low $\pT$ ($\pT<2$\,GeV/$c$) almost all the $\jpsi$ at midrapidity are produced via (re)generation, while this fraction reduces to $\sim$60\% at forward rapidity.  
The SHMc model describes very well the data at low transverse momentum and the centrality dependence with $\jpsi$ production entirely at the hadronization of the QGP. It underestimates $\raa$ at high $\pT$, as a consequence of including only the pp-like contribution from the corona region and no $\jpsi$ production in jets. 

Other models (not shown in \textbf{Figure~\ref{fig:exp-jpsi-pt_npart}}) are also available and fairly describe the data (see e.g.\,\cite{ALICE:2022wpn,ALICE:2023gco}). Alternative transport models, such as the one described in Ref.\,\cite{Zhou:2014kka}, employ slightly different implementations of the rate equation and open charm cross-section. Other approaches, such as the ``comover model''\,\cite{Ferreiro:2012rq} are based on the assumption that the modification of the charmonium yields is due to (re)generation and suppression mechanisms, with the suppression due to the interactions of the pre-resonant $\jpsi$ state with comovers of partonic or hadronic origin, produced in the same kinematical region. In this model, the comovers density is tuned on the measured hadron yields, the dissociation cross section is tuned on low energy data, assuming no $\sqrtsNN$ dependence and the (re)generation effects are based on a transport equation. 

The uncertainties of the models are still rather large when compared to the precision currently reached by the experimental data. This is mainly a consequence of the uncertainty on the inclusive charm cross-section for Pb–Pb collisions.
The equivalent value per nucleon-nucleon collision, including nuclear shadowing for Pb-Pb, is \dsigmacc = 0.68\,$\pm$\,0.10 mb and \dsigmacc= 0.93\,$\pm$\,0.12 mb for the SHMc and TAMU models, respectively, see Section\,\ref{sec:theoretical_models}.

In the bottom panels of \textbf{Figure\,\ref{fig:exp-jpsi-pt_npart}}, the inclusive $\jpsi$ $\pT$-differential yields, $\dd^{2}N/(\dd{\rm y}\dd p_{\rm T})$ in Pb-Pb collisions are also shown, in the two rapidity ranges under study, also in direct comparison with the theory calculations\,\cite{Andronic:2021erx,Wu:2024gil}.
The good agreement of the TAMU model and of the SHMc one, in this case in the low $\pT$ region, already observed for the $\raa$, is confirmed in the description of the spectra. We recall that the models genuinely predict yields, $\raa$ is obtained via a normalization to the pp measurement. The SHMc calculations shown in \textbf{Figure\,\ref{fig:exp-jpsi-pt_npart}} are based on a parametrized (blast wave) hydrodynamical flow\,\cite{Andronic:2021erx}, a calculation with a full hydrodynamical treatment\,\cite{Andronic:2023tui} leads to a harder $\pT$ spectrum (noticed earlier in Ref.\,\cite{He:2021zej}), consequently describing the data less well. It is shown in Ref.\,\cite{Andronic:2023tui} that a slightly narrower spatial distribution of (anti)charm quarks leads to a softer spectrum, in agreement with data. A more refined treatment in hydrodynamics, including diffusion of charm quarks and atiquarks, is performed in Refs.\,\cite{He:2021zej,Capellino:2023cxe}.

\subsubsection{$\jpsi$ elliptic flow}
In parallel with the extensive studies of the $\jpsi$ $\raa$, further insights on the behavior of charmonium in AA collisions can be inferred by measuring the $\jpsi$ elliptic flow, evaluated through the $\vtwo$ observable. 

\begin{figure}[htb]
\begin{tabular}{lcr} \begin{minipage}{.49\textwidth}
\includegraphics[width=1.\linewidth]{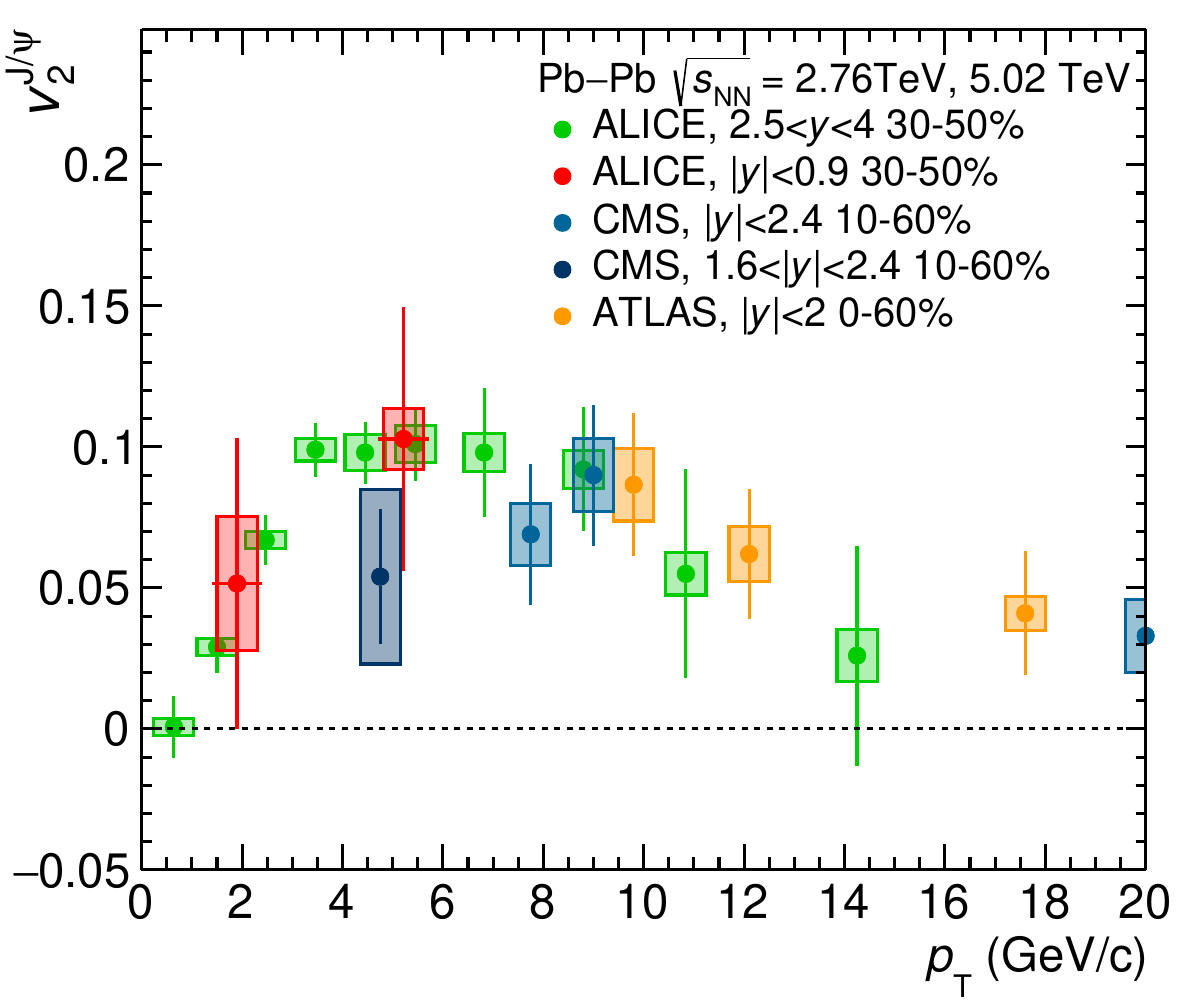}
\end{minipage} &\begin{minipage}{.49\textwidth}
\includegraphics[width=1.\linewidth]{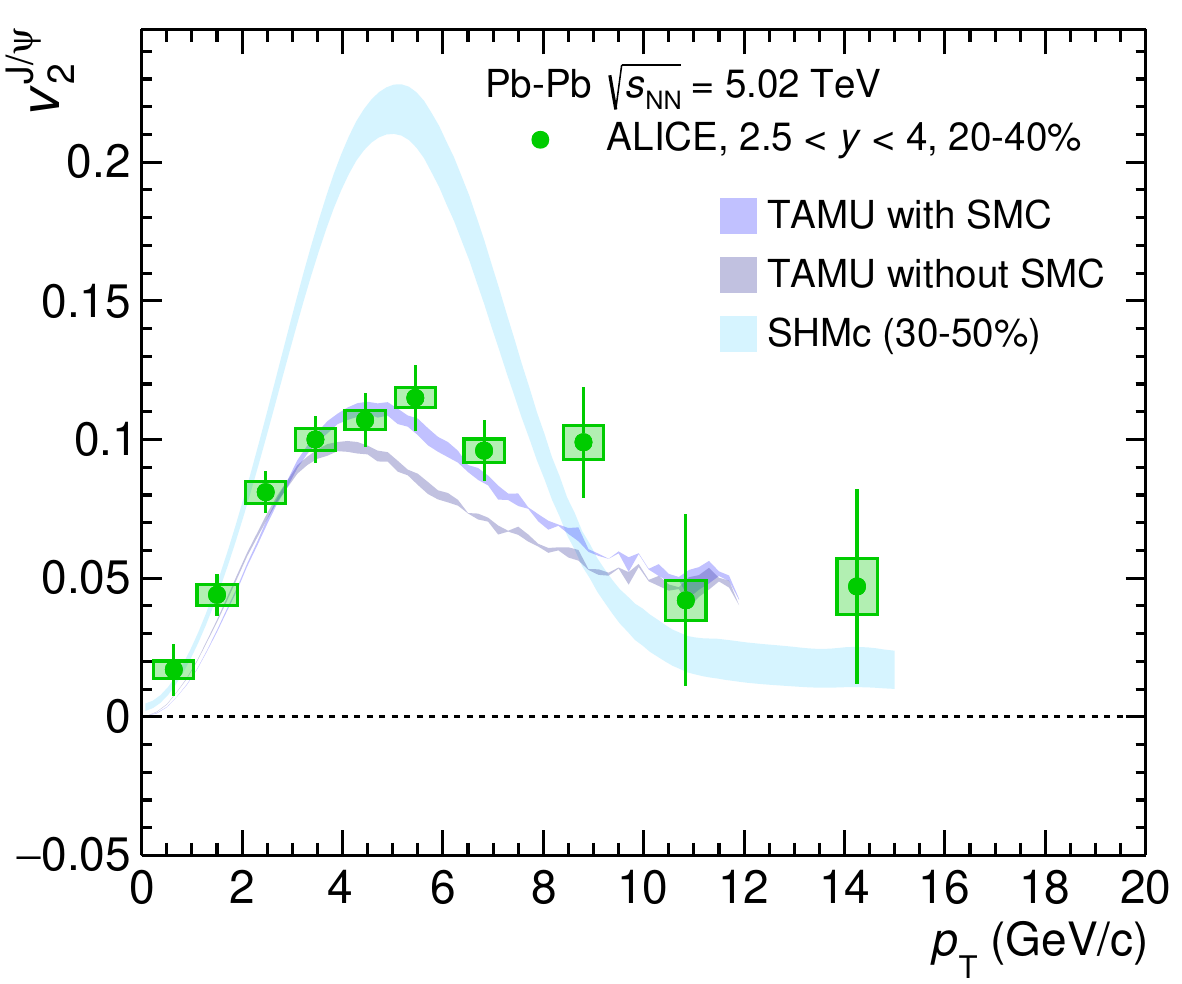}
\end{minipage} \end{tabular}
\caption{The $\pT$-dependence of $\jpsi$ $\vtwo$ is shown, compiling results from  LHC~\cite{ATLAS:2018xms,ALICE:2017quq,CMS:2023mtk} experiments (left panel). The $\pT$-dependence of the ALICE $\vtwo$ results at forward rapidity, in the centrality interval 20-40\%, is compared to theory calculations from the TAMU\cite{He:2021zej} and SHMc\,\cite{Andronic:2023tui} (right panel). We note that the SHMc calculation is for a slightly different centrality range, i.e. 30-50\%.}
\label{fig:exp-jpsi-flow}
\end{figure}

The compilation of the available $\vtwo$ results at LHC energies is shown in the left panel of \textbf{Figure~\ref{fig:exp-jpsi-flow}}. 
Although the measurements included in the plot have slightly different kinematic coverage or are at a different collision energy and centrality, several features can be observed. 
The $\jpsi$ $\vtwo$ shows a rise towards intermediate $\pT$, reaching a value of 0.1 in semi-central collisions and $\pT\sim 4$\,GeV/$c$. 
The behavior is explained assuming that a large fraction of the detected $\jpsi$ originates from the (re)generation from free charm quarks, which acquire their anisotropy by taking part in the collective expansion of the system. This observation confirms once more the interpretation of the $\raa$ measurements. It should be noted that at RHIC energies the $\jpsi$ elliptic flow is compatible with zero, albeit with large uncertainties~\cite{PHENIX:2024axj,STAR:2012jzy}.
In the right panel of \textbf{Figure~\ref{fig:exp-jpsi-flow}}, the ALICE $\vtwo$ results obtained in the rapidity region $2.5< y <4$, in 20-40\% centrality, are compared to the previously discussed theory models, TAMU\,\cite{He:2021zej} and SHMc\,\cite{Andronic:2023tui}. The TAMU model describes the $\vtwo$ over the fully explored range, thanks to the inclusion of space-momentum correlations of the diffusing charm and anti-charm quarks in the expanding fireball. The model describes very well the data also for the high-$\pT$ region, demonstrating that (re)generation is important even at $\pT\simeq$\,8\,GeV/$c$.
The SHMc model predicts a rise of the $\vtwo$ reaching a maximum around $\pT\sim$\,5\,GeV/$c$, however, it overestimates the data significantly, underscoring that a more compact spatial distribution of the charm quarks compared to the lighter ones, arising from delayed charm thermalization, plays a role. The TAMU calculations\,\cite{He:2019tik} exhibit comparably large $\vtwo$ values for the (re)generated $\jpsi$, but the inclusion of the primordial component, which dominates for $\pT>5$\,GeV/$c$, leads to the very good description of the data.
In the SHMc calculations, the $\vtwo$ values are determined by the interplay between the contributions from the hydrodynamic flow and the isotropic corona, the latter becoming significant for $\pT\gtrsim$ 6\,GeV/$c$. For large $\pT$ values $\vtwo$ is likely due to path-length effects~\cite{Noronha-Hostler:2016eow}.
The $\vtwo$ measurement performed by ALICE at midrapidity (and included in the left panel of \textbf{Figure~\ref{fig:exp-jpsi-flow}}) confirms the observations at forward rapidity, but the large uncertainties so far prevent more quantitative conclusions on how $\vtwo$ depends on rapidity.

\subsubsection{$\jpsi$ polarization}
Finally, additional information on the quarkonium behavior in the QGP might be inferred by polarization measurements. First analyses from the ALICE Collaboration~\cite{ALICE:2020iev} have computed the polarization parameters in the commonly adopted helicity and Collins-Soper reference frames, in the rapidity range 2.5 $< y <$ 4 in Pb--Pb collisions. The parameters $\lambda_{\theta}$, $\lambda_{\phi}$ and $\lambda_{\theta\phi}$, which define the degree of polarization, are found to be compatible with zero, reaching a maximum of about two standard
deviations at low $\pT$, for both reference frames.

The degree of polarization can also be sensitive to the strong magnetic field created in the high-energy nuclear collisions, as well as to vorticity effects in the QGP. 
To test this, the polarization is evaluated with respect to the event plane of the collision~\cite{ALICE:2022dyy}. A small transverse polarization is measured, with a significance reaching 3.9$\sigma$ at intermediate $\pT$ ($2 < \pT < 4$ GeV/c) and centrality (30-50\%). This observation is roughly in agreement with a similar measurement for light vector mesons~\cite{ALICE:2019aid}, consistent with a possible quark polarization in the presence of a rotating fluid. 
However, a quantitative understanding of how the polarization of quarkonium is influenced and reflects the QGP properties would still require, from the theory point of view, detailed calculations connecting the $\jpsi$ production to the QGP properties and, from the experimental point of view,  larger luminosities.

\subsubsection{$\psip$ in AA collisions}
To have a complete overview of the charmonium production, the understanding of the role of the excited states is mandatory because, given the larger mass, the smaller binding energy and larger radius, they can be affected in a different way by the suppression and (re)generation mechanisms. Furthermore, since about $30\%$ of the prompt $\jpsi$ are produced by the feed-down of the $\psip$ and $\chic$ states, the understanding of their fate in the QGP allows to better interpret the fate of the directly produced $\jpsi$. 

The measurement of the $\chic$ states is rather complicated, due to the difficulties in the reconstruction of their radiative decay $\chic \rightarrow \jpsi~\gamma$. First results on the $\chi_{c1}+\chi_{c2}$ production in pA collisions from the LHCb Collaboration have been recently released~\cite{LHCb:2023apa}, but no measurement in AA are so far available.
Additionally, the $\psip$ measurements are challenging due to its branching ratio in the dimuon channel being approximately 7.5 times lower compared to the $\jpsi$, as well as its production cross section in pp collisions at LHC energy being about six times smaller. 

Already at SPS energies, the NA50 experiment~\cite{NA50:2006yzz} had observed a stronger suppression of this resonance with respect to the $\jpsi$. Therefore it is important to investigate its behavior also at LHC, where competitive mechanisms such as (re)generation might be involved. While the $\psip$ large radius and small binding energy favor its suppression with respect to the tightly bound $\jpsi$, the impact on the (re)generation processes on the $\psip$ is less clear. According to the TAMU model~\cite{Wu:2024gil}, for example, the (re)generation of $\jpsi$ and $\psip$ is sequential, with the $\psip$ being formed by (re)generation at lower temperatures than the $\jpsi$, with significant contributions not only from the QGP phase, but also from the hadronic phase.
In the SHMc model both $\psip$ and $\jpsi$ are produced exclusively at hadronization at the QCD crossover boundary. Consequently, their relative yield is determined solely by the mass difference and the temperature.

Before moving to the discussion of Pb--Pb data, it should be pointed out that the $\psip$ production shows already medium modifications in p--Pb and d--Au collisions, in particular in the backward rapidity region~\cite{ALICE:2014cgk,ALICE:2020vjy,LHCb:2016vqr,PHENIX:2016vmz}. Initial-state cold nuclear matter effects, such as shadowing, affect in a very similar way the $\jpsi$ and the $\psip$, being directly connected with the creation of the heavy-quark pair. But while the modification of the $\jpsi$ yields can be described just with these initial-state CNM effects, in the $\psip$ case additional final-state effects, such as interactions in a dense medium (of hadronic or partonic origin~\cite{Ma:2017rsu,Ferreiro:2014bia}) are needed to describe the data. This means that the medium created in p(d)-A collisions might already impact the yields of the loosely bound $\psip$.

\begin{figure}[hbt]
\begin{tabular}{lr} \begin{minipage}{.49\textwidth}
\includegraphics[width=1.\linewidth]{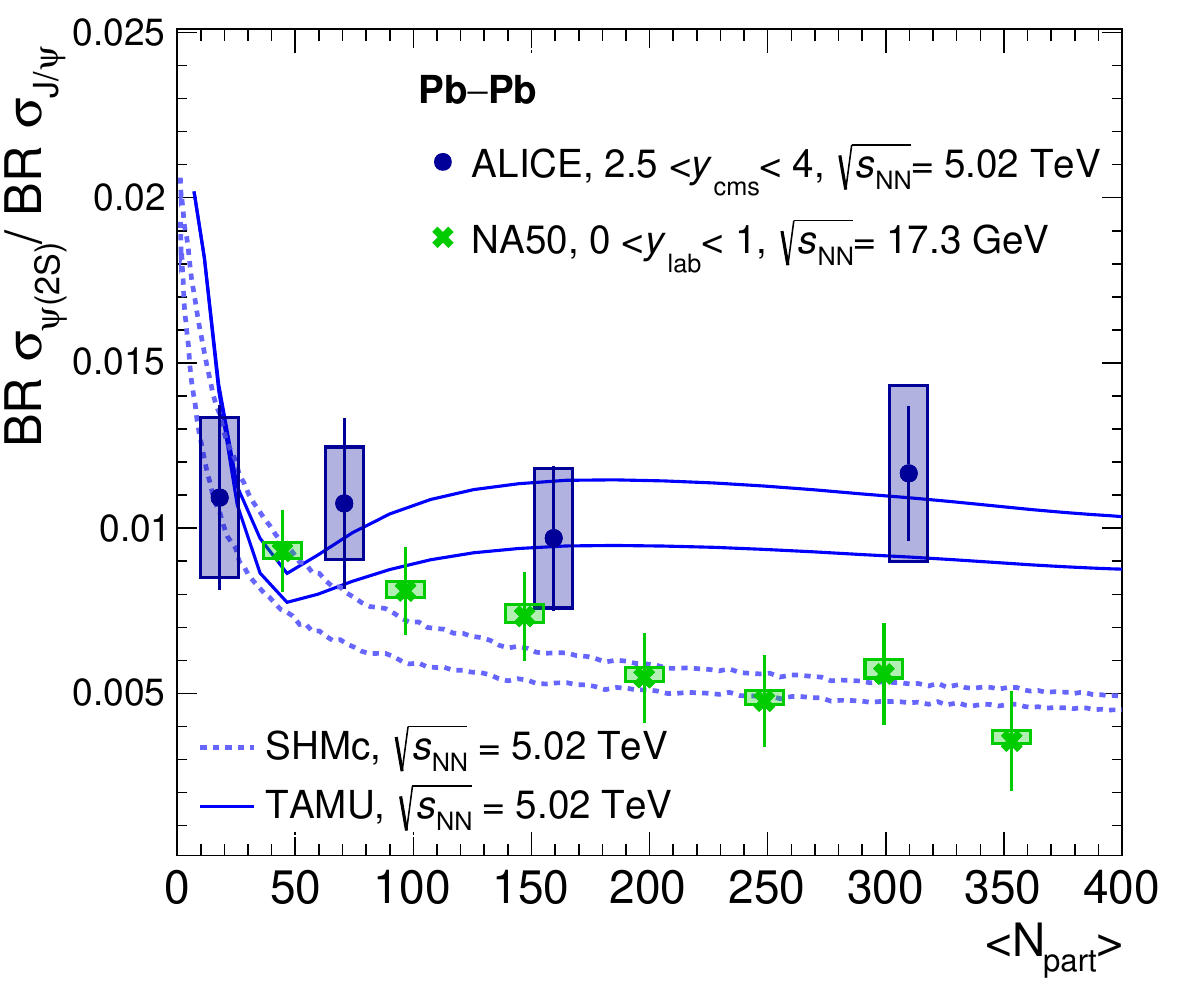}
\end{minipage} &\begin{minipage}{.49\textwidth}
\includegraphics[width=1.\linewidth]{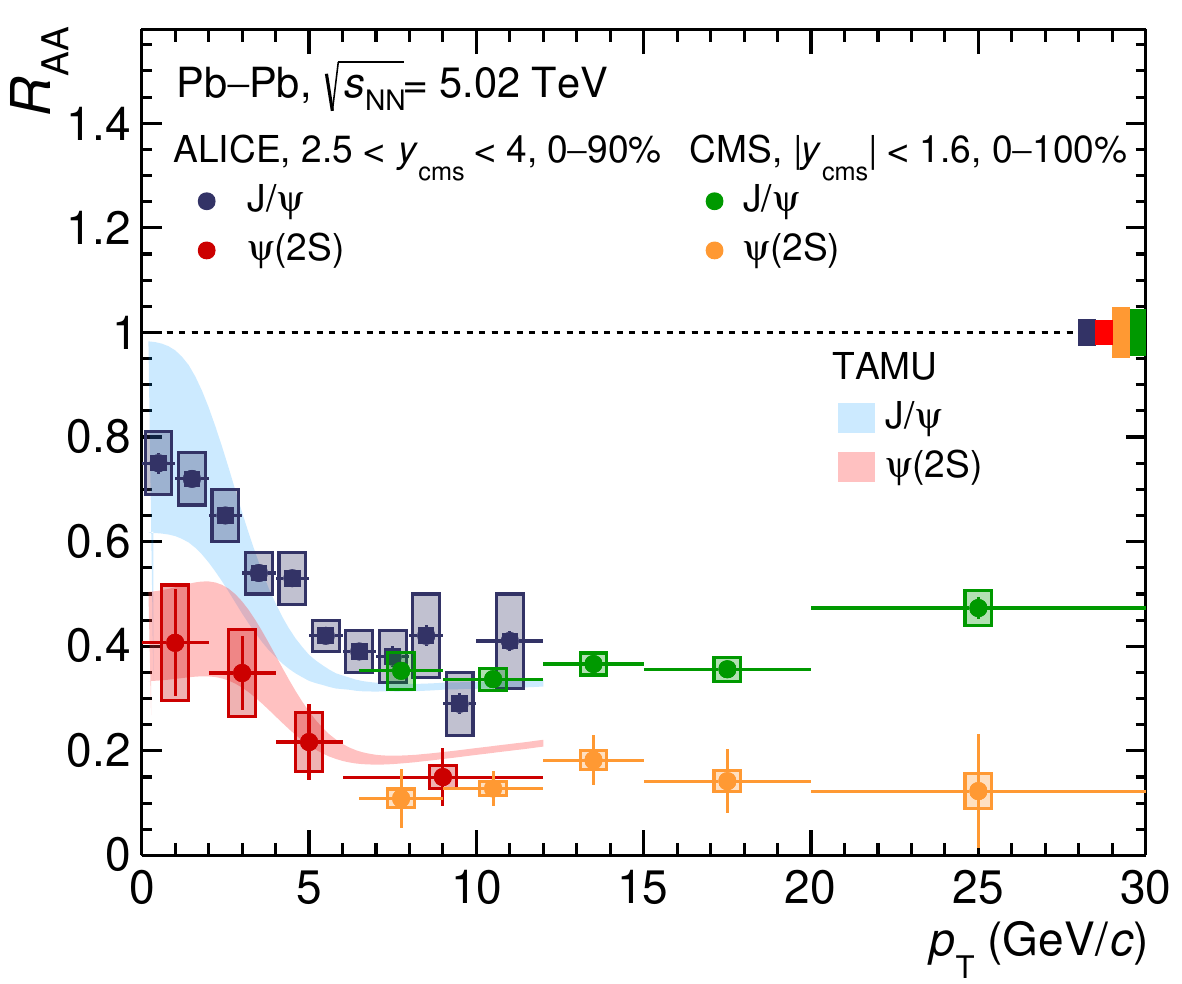}
\end{minipage} \end{tabular}
\caption{The ratio of the inclusive production cross sections of the $\psip$ to $\jpsi$, not corrected for the branching ratios in the dimuon channels, is plotted as a function of $\langle N_{\rm{part}}\rangle$. Results from ALICE~\cite{ALICE:2022jeh} and NA50~\cite{NA50:2006yzz}, in Pb--Pb collisions, are shown together with the theoretical curves from TAMU~\cite{Du:2015wha} and SHMc~\cite{Andronic:2017pug,Andronic:2019wva} at $\sqrtsNN$\,=\,5.02\,TeV (left panel). 
A compilation of $\psip$ and $\jpsi$ $\raa$ results is shown as a function of $\pT$~\cite{CMS:2017uuv,ALICE:2022jeh} (right panel). The TAMU~\cite{Wu:2024gil} calculations are also included for both resonances. It should be noted that ALICE results refer to inclusive charmonia, while the CMS ones refer to prompt production.} 
\label{fig:exp-psip2j}
\end{figure}

In \textbf{Figure~\ref{fig:exp-psip2j}} (left panel) the relative production of $\psip$ and $\jpsi$ is shown as a function of $\langle N_{\rm{part}}\rangle$. ALICE results~\cite{ALICE:2022jeh} in the region $2.5< y <4$, even if limited to four centrality bins, do not show a significant centrality dependence. The NA50~\cite{NA50:2006yzz} results at the SPS, on the contrary, do show a significant decrease versus centrality, being more suppressed in the most central collisions. As pointed out in Ref.~\cite{ALICE:2022jeh} part of the different behavior observed in the two results could be due to the different size of the non-prompt component, almost negligible at low energy. The TAMU~\cite{Du:2015wha} calculation describes, within the uncertainties, the ALICE observed trend, while the SHMc~\cite{Andronic:2017pug,Andronic:2019wva} one slightly underestimates the data in the most central collisions. We note that for the SHMc model the predictions are identical for the LHC and SPS energies, since, as already discussed, they depend on the hadronization temperature, which was observed to not change from SPS to LHC energies\,\cite{Andronic:2017pug}. The model describes the SPS data very well. 
The centrality dependence of the ratio is in the SHMc determined by the superposition of the core and corona components.

Also the $\psip$, as the $\jpsi$, can be studied by the ALICE experiment down to zero transverse momentum, as visible in the right panel of \textbf{Figure~\ref{fig:exp-psip2j}}, where a compilation of the $\raa$ of the two mesons is shown as a function of $\pT$~\cite{CMS:2017uuv,ALICE:2022jeh}.
The availability of results from both the ALICE and the CMS experiments allows a detailed study over an extended $\pT$ range, from zero up to 30 GeV/$c$. 
Several features are visible. First of all, the $\psip$ and the $\jpsi$ $\raa$ follow a very similar trend, but the $\psip$ shows a stronger suppression, by almost a factor of two, over the explored $\pT$ range. A clear hierarchy in the $\raa$ is hence visible. Furthermore, in the low $\pT$ region, also the $\psip$ $\raa$ has a rise towards low $\pT$, similar to the one observed in the $\jpsi$, and interpreted as due to (re)generation mechanism. This observation hints at the conclusion that similar processes are at play also in the $\psip$ case. 
A comparison with the predictions of the TAMU model \cite{Wu:2024gil} is also shown in the same plot. The model, which was already successfully describing the $\jpsi$, nicely describes the $\psip$ $\raa$, pointing out that, also in the case of this loosely bound resonance, suppression and (re)generation mechanism play a role. According to Ref.~\cite{Wu:2024gil}, the majority of the $\psip$ are produced by (re)generation at low $\pT$ region, in central and semi-central collisions. The contribution of primordial $\psip$ is relevant only for $\pT>5$\,GeV/$c$, where the production by (re)generation practically vanishes. 
  
%%% Upsilon %%%%%%%%%%
\subsection{Bottomonium experimental results}
%%%%%%%%%%%%%%%%%%%%%%%
Quarkonium studies received a further significant boost when the first bottomonium measurements became available at the LHC. 
Together with the observation of the $\jpsi$ $\raa$ increase towards low $\pT$ proving the production via (re)generation, the eagerly awaited observation, done by the CMS collaboration, of the suppression of $\Upsilon$ mesons in Pb--Pb collisions can be considered as a breakthrough~\cite{CMS:2012gvv}. The first results were soon confirmed by the more precise and differential measurements of $\raa$ for all the $\Upsilon$ states by CMS \cite{CMS:2023lfu,CMS:2017ycw,CMS:2018zza} and ATLAS \cite{ATLAS:2022exb} at midrapidity and ALICE at forward rapidity \cite{ALICE:2020wwx}. An ordering in the suppression of the three states is observed, with the radially-excited states $\YiiS$ and $\YiiiS$ significantly more suppressed compared to $\YiS$.

\begin{figure}[htb]
\begin{tabular}{lr} \begin{minipage}{.49\textwidth}
\includegraphics[width=1.\linewidth]{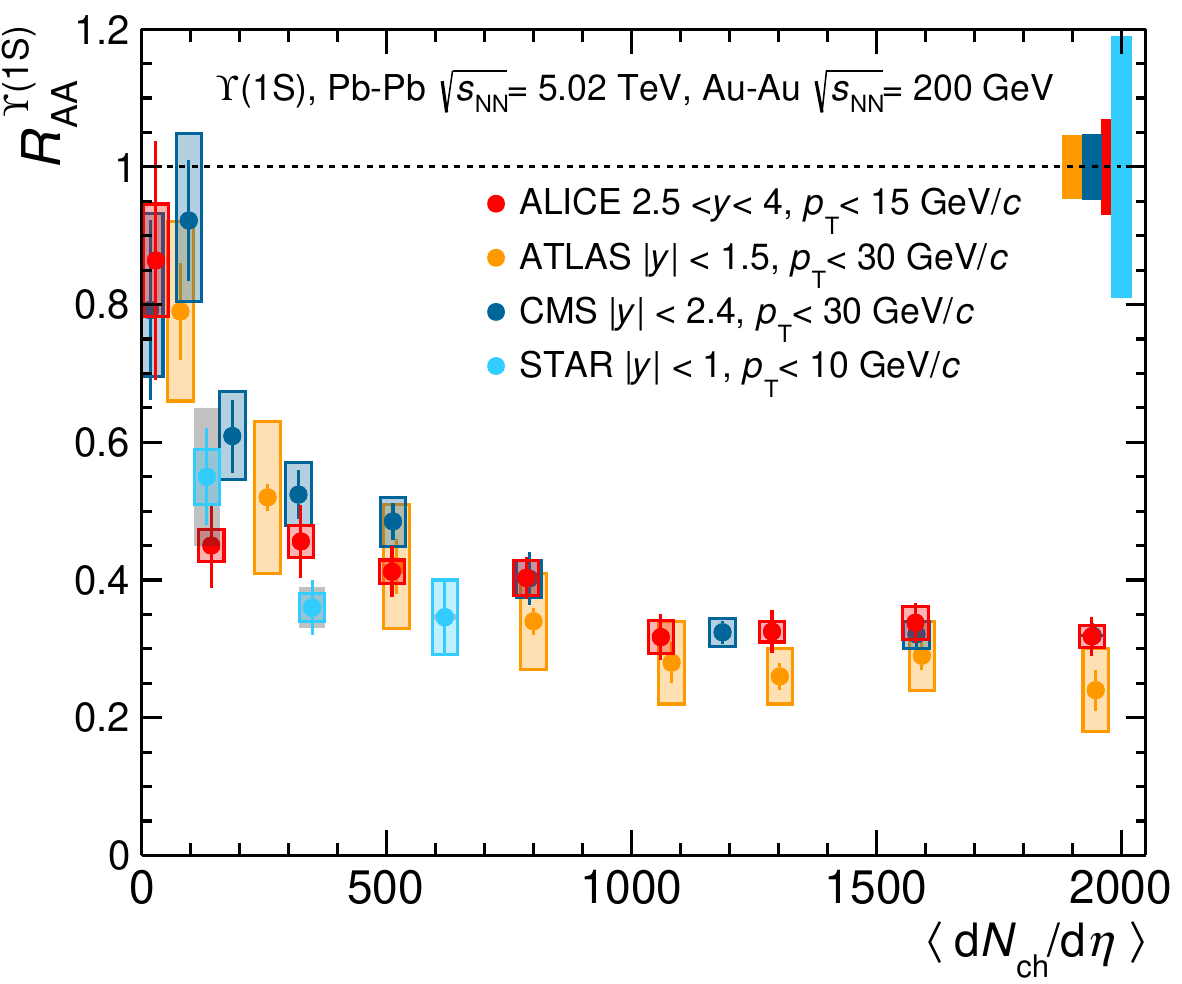}
\end{minipage} &\begin{minipage}{.49\textwidth}
\includegraphics[width=1.\linewidth]{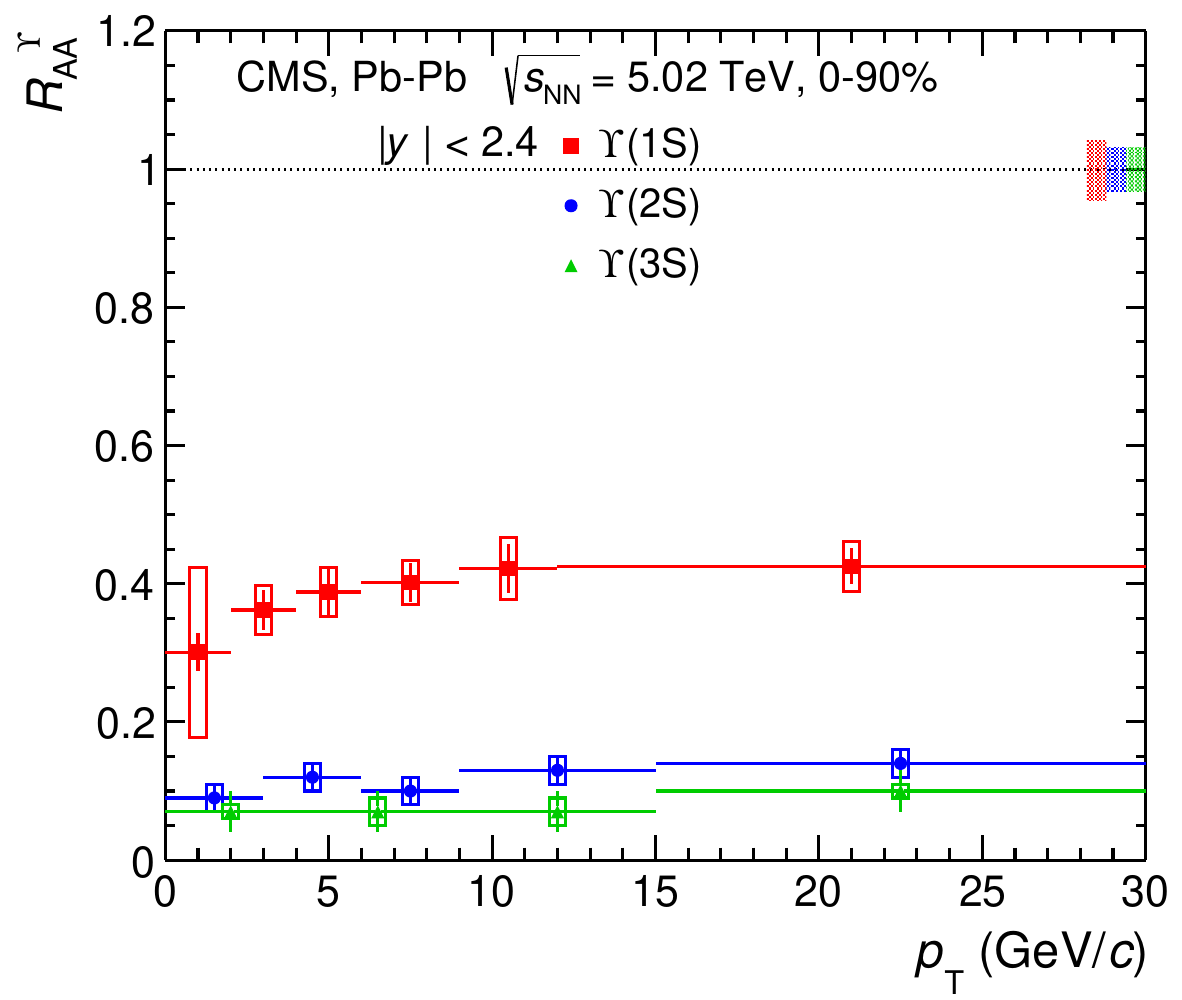}
\end{minipage} \\
\begin{minipage}{.49\textwidth}
\includegraphics[width=1.\linewidth]{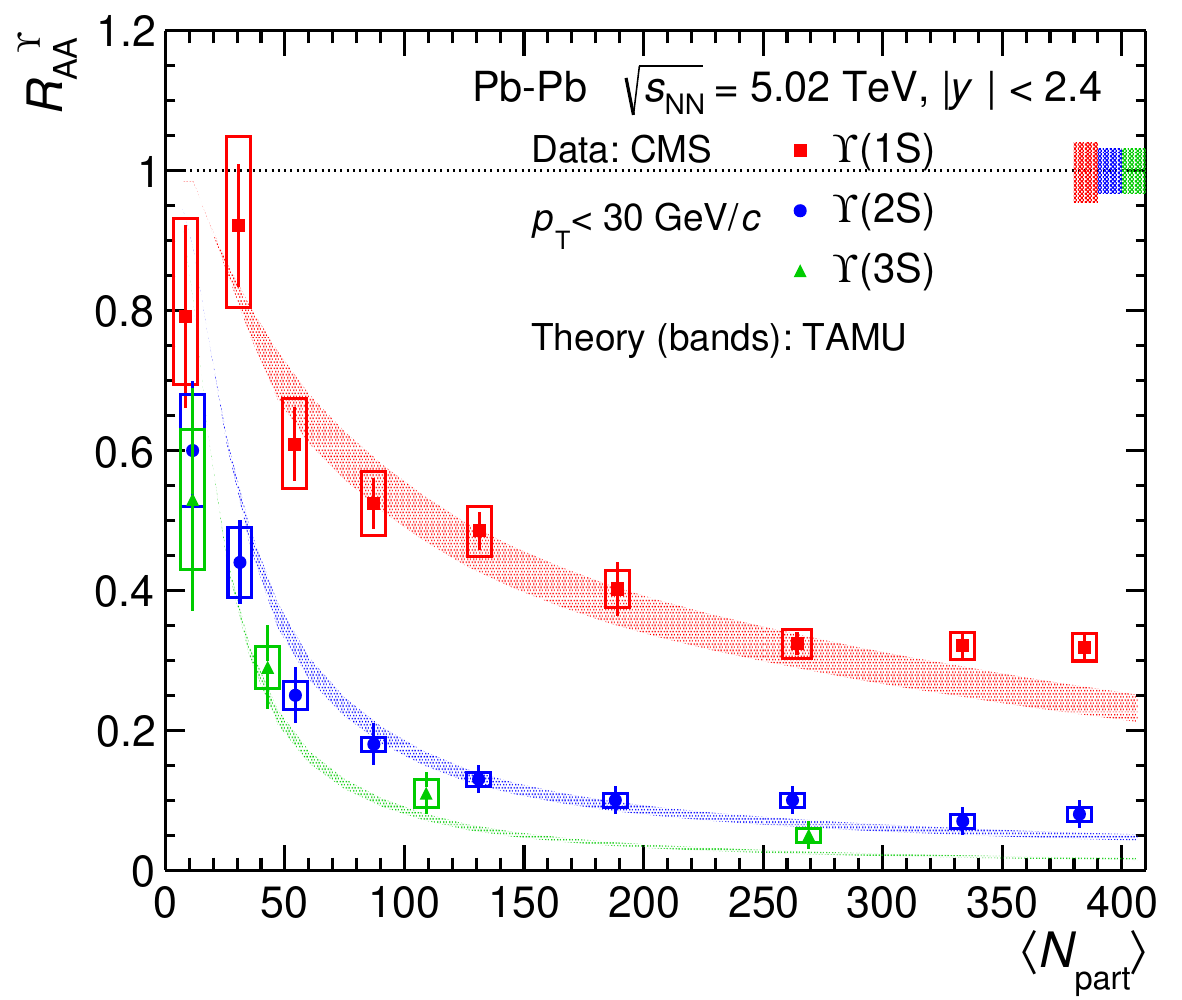}
\end{minipage} &\begin{minipage}{.49\textwidth}
\includegraphics[width=1.\linewidth]{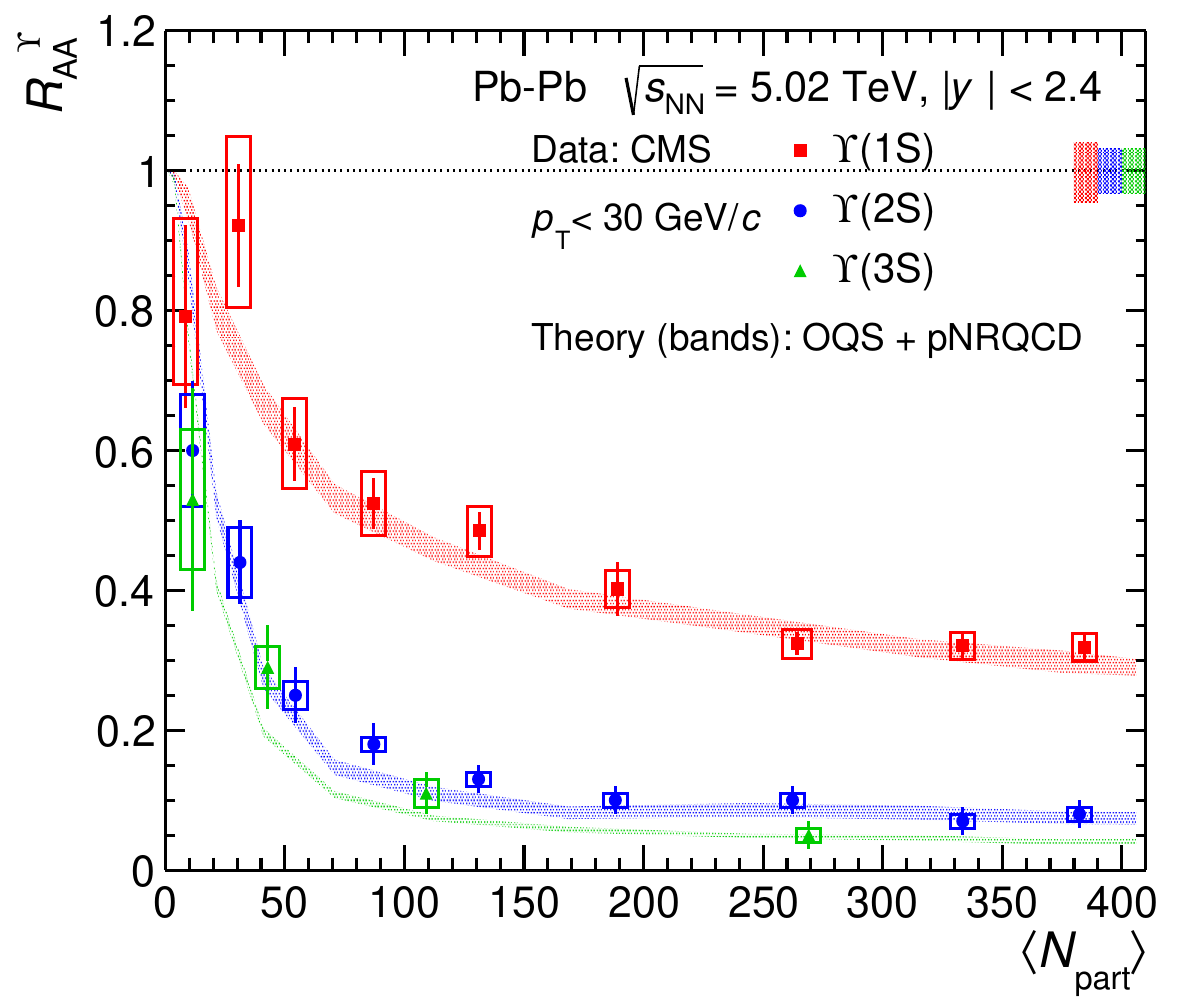}
\end{minipage} \end{tabular}
\caption{The  $\YiS$ $\raa$ dependence on $\langle \dd N_{ch}/\dd\eta \rangle$ for Au-Au collisions at $\sqrtsNN= 200$~GeV~\cite{STAR:2022rpk} and Pb--Pb at $\sqrtsNN= 5.02$~TeV\,\cite{CMS:2023lfu,CMS:2017ycw,CMS:2018zza,ATLAS:2022exb,ALICE:2020wwx} is shown (top left panel). The $\raa$ of the three $\Upsilon$(nS) states as a function of $\langle N_{\rm{part}}\rangle$, and compared to calculations from TAMU \,\cite{Du:2017qkv} and the OQS+pNRQCD approach of the Munich-KSU group\,\cite{Brambilla:2023hkw}, is shown in the bottom left and right panels, respectively, while the $\pT$ dependence~\cite{CMS:2023lfu} is shown in the top right panel. }
\label{fig:exp-y-npart}
 \end{figure}

A compilation of the $\YiS$ $\raa$ results is presented in the top left panel of \textbf{Figure~\ref{fig:exp-y-npart}}, where all the available measurements, at several center-of-mass energies are shown as a function of the charged particle multiplicity evaluated at midrapidity. When plotted versus $\langle\mathrm{d}N/\mathrm{d}\eta\rangle$, contrarily to what happens for charmonia, all the results follow the same trend, with the suppression magnitude and pattern similar at LHC~\cite{CMS:2023lfu,CMS:2017ycw,CMS:2018zza,ATLAS:2022exb,ALICE:2020wwx} and at RHIC~\cite{STAR:2022rpk} energies. Another feature visible in the $\YiS$ $\raa$ pattern is a significant reduction from the peripheral to the most central events, by almost a factor 4 at LHC. 

The behavior of the $\YS$ states is clearly different from the charmonium one. 
The $\YiS$ suppression does not exhibit a significant energy dependence. It also presents a similar magnitude at forward rapidity and midrapidity and has a mild $\pT$ dependence, as shown in the four panels of \textbf{Figure~\ref{fig:exp-y-npart}}. All these features point to a different role of the (re)generation mechanism when $\mathrm{b}$ quarks are involved. The expected number of $\bbbar$ pairs in central Pb--Pb collisions is about 4 in the full phase space, with about 0.6 pairs per unit of rapidity at midrapidity and is smaller by about a factor of two for the range $2.5<y<4$. 

When comparing the $\raa$ of the three $\YS$, the suppression is gradually stronger going from the 1S to the 2S and 3S states. This behavior can be observed both as a function of centrality (bottom panels of \textbf{Figure~\ref{fig:exp-y-npart}}) and as a function of the transverse momentum (top right panel). In most central collisions, the $\YiS$ are suppressed by a factor $\sim$3, the $\YiiS$ by a factor $\sim$10 and the $\YiiiS$ are almost vanished.  This striking pattern is denoted as ``sequential suppression''. 
It should be noted that, since in pp collisions, $\sim$25\% of the measured $\YiS$ yield results from the feed-down from other states, a significant amount of $\YiS$ suppression may arise as a consequence of the melting of the excited states in the QGP. 
Together with the impact of the excited states feed-down, also cold nuclear matter effects should be taken into account when interpreting the $\Upsilon(\mathrm{nS})$ modification in AA collisions. Proton-nucleus results~\cite{ALICE:2019qie,ATLAS:2017prf,LHCb:2018psc,CMS:2013jsu} show that effects such as shadowing, affect bottomonia, leading to a reduction of their yields, even if less than for the $\jpsi$. A precise measurement of the size of these effects is hence relevant to understand if even the directly produced, and strongly bound, $\YiS$ melts in the QGP.

The comparison of bottomonium results to two theoretical models is shown in the bottom panels of \textbf{Figure~\ref{fig:exp-y-npart}}. It can be noticed that the models nicely describe the measurements for the whole bottomonium family. The TAMU calculations\,\cite{Du:2017qkv} include regeneration, which leads to a contribution, for instance, of about 20\% for the 1S state and about 40\% for the 2S state, dependent on centrality and $\pT$.
It is worth noting that the model achieved a very good description of RHIC data at $\sqrtsNN=200$\,GeV, in this case with a significantly lower amount of (re)generation\,\cite{Du:2017qkv}.
The comparison to the OQS+pNRQCD calculations\,\cite{Brambilla:2023hkw} is also shown. In this case as well, the inclusion of (re)generation, of a smaller amount compared to the TAMU case, improves the agreement with the data.  
Predictions within the SHM \cite{Andronic:2022ucg} could describe the data, but assuming that only 30-50\% of bottom-quark pairs are thermalized and consequently participate in the hadronization at the QGP crossover boundary.

The data are shown as a function of $\pT$ in the top right panel of \textbf{Figure~\ref{fig:exp-y-npart}} for midrapidity in 0-90\% Pb-Pb collisions. Also in this case, the observed trend is rather different compared to the charmonium one. No prominent features were observed, except a small increase of $\raa$ vs. $\pT$ for the $\YiS$ state and the sequential suppression observed as a function of centrality is confirmed. 
The theory calculations which describe the centrality dependence achieve a good description also of the $\pT$ dependence of $\raa$ (not shown here, see Ref.\,\cite{CMS:2023lfu}).

The elliptic flow of $\YiS$ and $\YiiS$ mesons was measured at LHC energies. Even if the uncertainties are still large, the $\vtwo$ turns out to be compatible with zero, both at forward rapidity~\cite{Acharya:2019hlv} and at midrapidity~\cite{Sirunyan:2020qec}.

\section{WRAP UP ON CURRENT UNDERSTANDING OF QUARKONIA IN QGP}
\label{sec:wrapup}
Quarkonium is certainly one of the most prominent probes for QGP studies, its production being sensitive to the medium created in the collision. One of the quarkonium points of strength is that it exists in a large variety of states, characterized by very different binding energies, from $\sim$ 50 MeV for the $\psip$ up to $\sim$ 1.1 GeV for the $\YiS$. All these states are expected to be sensitive in a different way to the created medium, hence providing handles to understand the QGP properties.
Furthermore, many measurements are available over a broad range of collision energies, 
from $\sqrtsNN$ =~17 GeV at SPS up to $\sqrtsNN$ =~5.36 TeV at LHC and a high level of precision is now reached. 

The most striking feature is that the $\jpsi$ behaviour at LHC energies is very different from the one at lower energies. At LHC, the $\pT$ dependence of the $\raa$ is rather strong, with $\raa$ even slightly exceeding unity at low $\pT$ and midrapidity followed by a strong decrease towards high $\pT$. On the contrary, at lower energies the $\pT$ dependence is rather mild.
Furthermore, at LHC, the $\jpsi$ presents also a $\vtwo$ significantly different from zero, suggesting that charm quarks thermalize in the medium and take part in the collective flow.

Theory models such as the statistical hadronization or those based on transport equations provide a quantitative description of the $\raa$ and $\vtwo$ results assuming an interplay of suppression and (re)generation mechanisms. As observed, at LHC the (re)generation process dominates for $\jpsi$ produced at midrapidity and low $\pT$, since this is the kinematic region where the bulk of the charm quark production is. The theoretical description of the data, with about 16 $\ccbar$ pairs in a QGP volume of about 5000\,fm$^3$ in central Pb-Pb collisions at midrapidity, provides a clear demonstration of deconfinement, with charmonium production predominantly at or close to the QCD crossover boundary.

In the bottomonium sector, a very clear hierarchy in the suppression of the $\Upsilon(\mathrm{nS})$ resonances is observed, with the $\YiS$ $\raa$ reaching 0.4 in most central collisions, while the $\raa$ of the $\YiiS$ reaches 0.1 and the $\YiiiS$ is almost vanished.
Several theoretical models describe these measurements, differing in the way they realize destruction and (re)generation of the $\YS$ states. Remarkable new developments include the quantum treatment of the processes, where new insights are still expected. For bottomonium, the suppression mechanism is the dominant one, even if a small amount of (re)generation might also be present. 

\begin{figure}[htb]
\begin{tabular}{lr} \begin{minipage}{.49\textwidth}
\includegraphics[width=1.\linewidth]{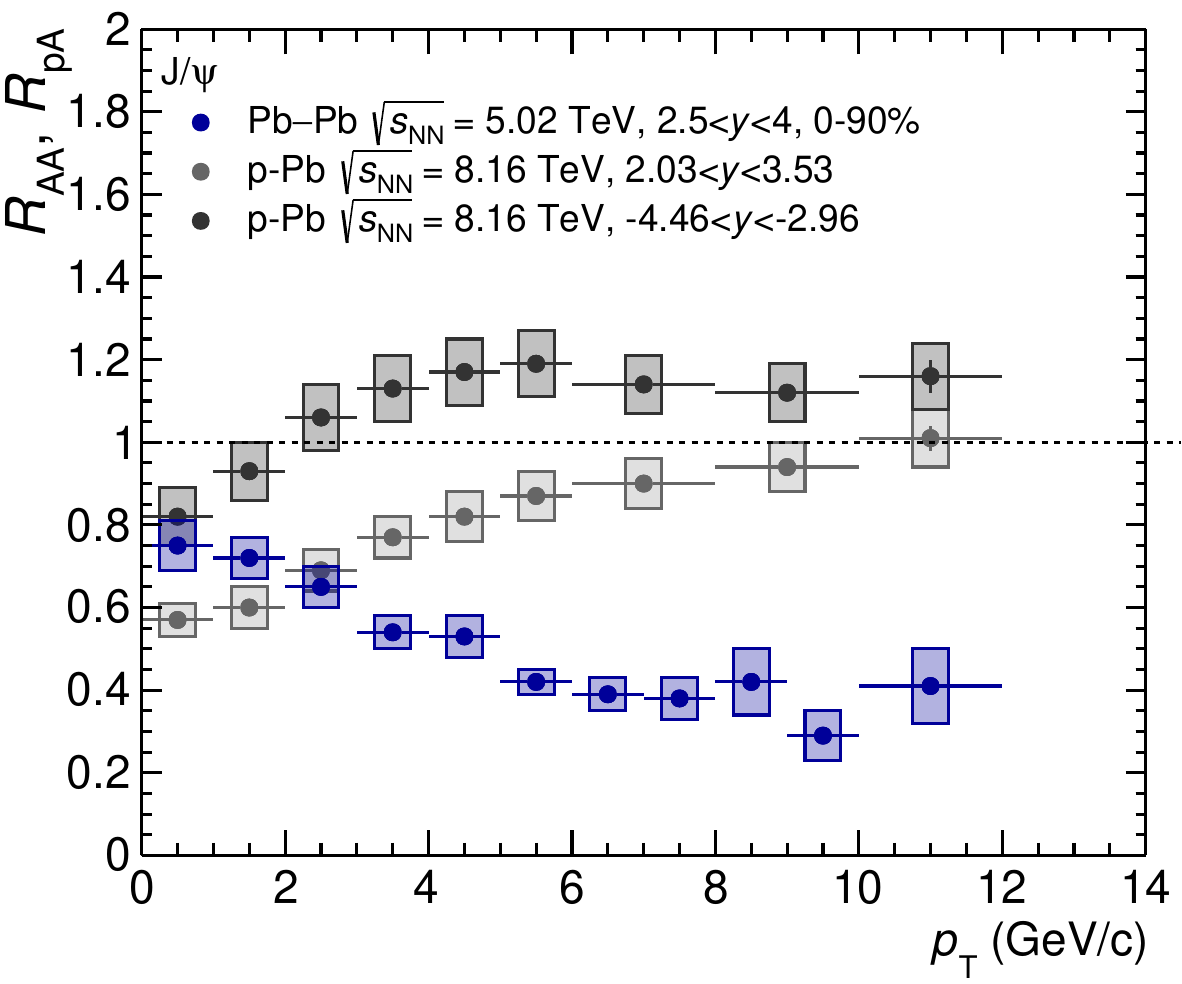}
\end{minipage} &\begin{minipage}{.49\textwidth}
\includegraphics[width=1.\linewidth]{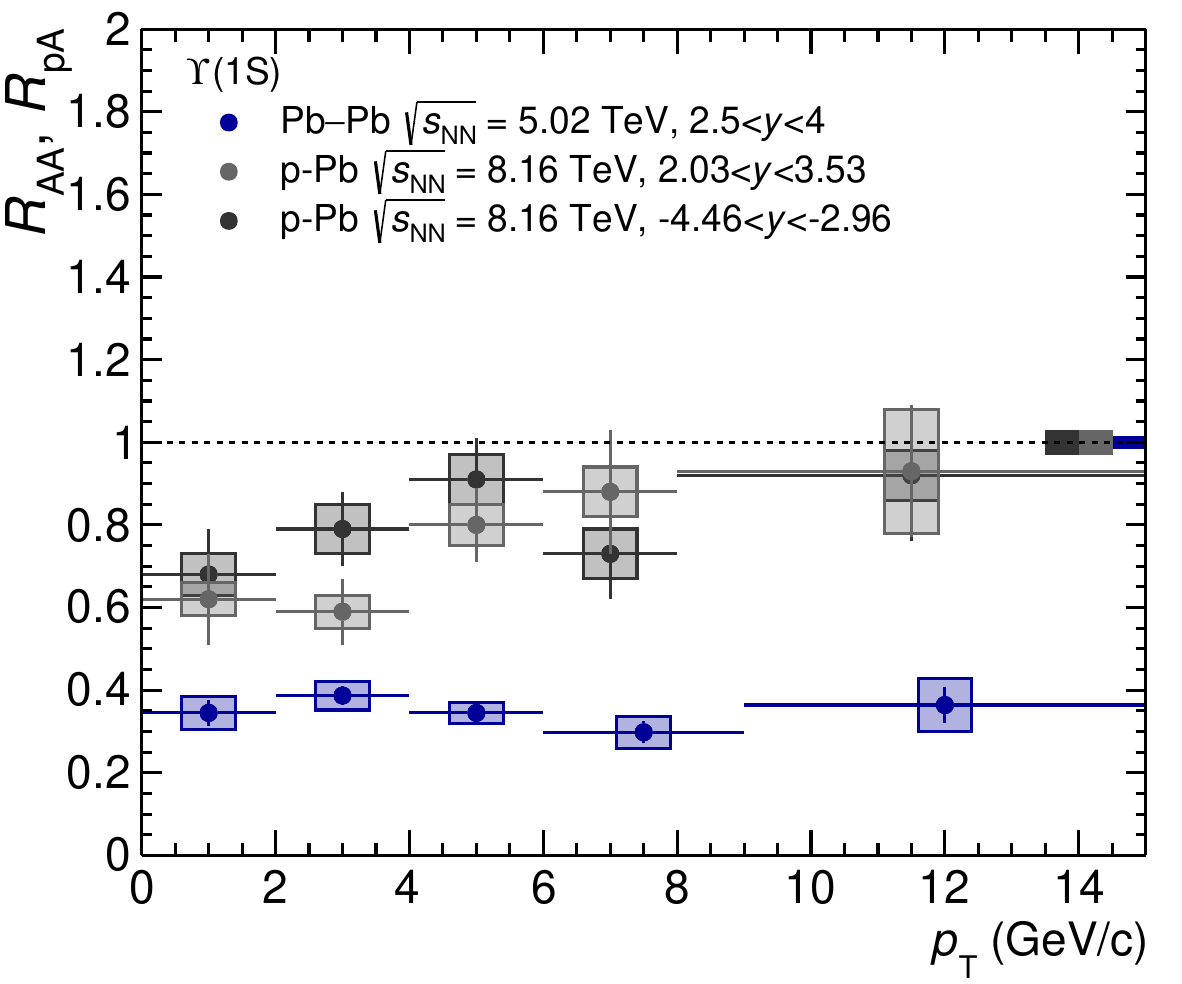}
\end{minipage} \end{tabular}
\caption{$\jpsi$~\cite{ALICE:2019lga} and $\YiS$~\cite{ALICE:2020wwx} $\raa$ as a function of $\pT$ (left and right panels, respectively), measured at forward rapidity in Pb--Pb collisions at $\sqrtsNN$ = 5.02 TeV (blue symbols). Results are compared to the corresponding $R_{pA}$ measured in pA collisions~\cite{ALICE:2019qie,ALICE:2018mml} (gray symbols).}
\label{fig:pA_AA}
\end{figure}

As a final wrap-up, we can compare the $\jpsi$ and $\YiS$ $\raa$ results obtained in Pb--Pb collisions at LHC, with the corresponding measurements in p-Pb. The $\raa$ as a function of $\pT$, as obtained by the ALICE experiment, is shown in \textbf{Figure~\ref{fig:pA_AA}}.
The pA data have been collected in two different rapidity ranges ($2.03 < y < 3.53$ and $-4.46 < y < -2.96$), corresponding to the proton or the Pb ions going towards the ALICE muon spectrometer, respectively. Despite the slightly different collision energy and kinematic range, a striking difference between the $\pT$ dependences in pA and AA can be observed. In particular, at high $\pT$, the $R_{pA}$ approach unity, suggesting a vanishing of modification effects, while in Pb--Pb $\raa$ is  significantly reduced. This is a clear indication that, in the two systems, different mechanisms are at play, both for the $\jpsi$ and for the $\YiS$.   

Going one step further, a more quantitative evaluation of the impact of CNM effects on Pb--Pb data can be performed, based on the reasonable assumption that, as discussed in  Ref.~\cite{ALICE:2019qie,ALICE:2013snh,ALICE:2015sru} and Section~\ref{sec:pppA}, cold nuclear matter effects, and in particular shadowing, are enough to describe the pA data for quarkonium ground states. The Bjorken-x ranges probed by the $\jpsi$ production process in the two colliding Pb nuclei, assuming a $gg \rightarrow \jpsi$ mechanism, are only slightly shifted
compared to the corresponding intervals for p--Pb and Pb--p, despite the slightly different center-of-mass energy and kinematic coverage (see  Ref.~\cite{ALICE:2013snh,ALICE:2015sru}) . Hence, assuming shadowing as the  dominant contribution and that CNM effects on the two colliding nuclei are factorizable, the shadowing contribution in Pb--Pb collisions can be evaluated as the product 
$R_{pA} \times R_{Ap}$~\cite{ALICE:2013snh,ALICE:2015sru}.
As shown in Ref.~\cite{ALICE:2022wpn}, CNM effects in Pb--Pb collisions can then be canceled out by studying the ratio $S_{\jpsi} = \raa / (R_{pA} \times R_{Ap})$. $S_{\jpsi}$ shows a significant $\pT$ dependence: at high $\pT$ a strong suppression is visible, with $S_{\jpsi}$ values reaching $\sim0.3$, while, at low $\pT$, $S_{\jpsi}$ exceeds unity.

Also for $\YiS$~\cite{ALICE:2022wpn} the $S_{\YiS}$ shows a strong $\pT$ dependence, reaching the same level as the $S_{\jpsi}$ at high $\pT$, while at low $\pT$ it approaches unity. Again, this behavior confirms that the role of (re)generation for bottomonia is less relevant than for charmonia. 
However, to precisely quantify the suppression of the direct $\YiS$ a precise knowledge of both the feed-down values and their $\pT$ dependence, and the suppression for all the excited states ($\YiiS$, $\YiiiS$, and all the $\chi_b$ states) is needed. While $\raa$ results for the $\Upsilon$ states are available, no measurements exist for the $\chi_b$. A numerical evaluation carried on in~\cite{ALICE:2022wpn} suggests that at low $\pT$ shadowing and feed-down effects could be responsible for most of the observed $\YiS$ suppression, while at high $\pT$, where the impact of the feed-down contribution gets larger and the CNM effects get weaker, there might be room for direct $\YiS$ suppression, but uncertainties prevent strong conclusions. 

\section{WHAT'S NEXT?}
Since the first measurements of  quarkonium in heavy-ion collisions, many steps forward in the understanding of its behavior have been taken, thanks to increasingly-precise sets of data collected over a broad $\sqrtsNN$ range, from the low SPS energies up to the top LHC ones, and to the wide kinematic region now covered, combining the results from different experiments. Furthermore, while the first quarkonium measurements were limited just to the $\jpsi$ and $\psip$, now also bottomonium  is extensively studied.  
However, there are still several aspects that remain to be clarified.

At LHC energies, the charmonium P-state, the $\chic$, has not yet been studied in AA collisions, given the extreme difficulty in the reconstruction of its radiative decay. The interest in the $\chic$ behavior is twofold. On one side, having an intermediate binding energy between the $\jpsi$ and the $\psip$ it would be interesting to understand how it is affected by a deconfined medium. 
On the other side, the $\chic$ feed-down contribution to the $\jpsi$ ranges between 15 and 30\%~\cite{Lansberg:2019adr} depending on the $\pT$. The measurement of the P-states should, hence,  provide additional insights on the in-medium modification of the direct $\jpsi$.  

Similarly, a precise assessment of the bottomonium excited states and their feed-down fractions would help in understanding if the observed $\YiS$ suppression is compatible with the suppression of the direct $\YiS$. 

Furthermore, even if not addressed in this review, in the past few years new lines of study opened up. First measurements of exotic states, such as the $\chi_{c1}(3872)$\footnote{$\chi_{c1}(3872)$, also denoted X(3872), is an exotic particle first observed by the Belle Collaboration~\cite{Belle:2003nnu}. Its quantum numbers are $J^{PC} = 1^{++}$ ~\cite{CDF:2006ocq,LHCb:2013kgk}, but its true nature remains unknown. Proposed hypotheses (e.g.  Ref.~\cite{Tornqvist:2004qy,Maiani:2004vq,Hanhart:2011jz}) range from it being a charmonium state, a tetraquark state or a $\mathrm{D}^{*}(2010)^{0}$ $\bar{\mathrm{D}^0}$ molecule.}  have been carried on in pA~\cite{LHCb:2024bpb} and AA collisions~\cite{CMS:2021znk}. The study of this state in the hot and dense matter can provide information on its still unknown nature and can shed further light on the role of (re)generation and suppression mechanisms. Current measurements are still limited by large uncertainties, preventing an answer, but the $\chi_{c1}(3872)$ studies will play a pivotal role in the physics program of the LHC experiments in the next years.

Exciting results are expected also from new experiments now starting to release the first results or which are proposed for the next years.
At RHIC energies, high precision data are expected from the sPHENIX experiment, the most recent among the heavy-ion experiments, which started commissioning and data taking only in 2023. sPHENIX aims to collect high luminosity Au-Au data at top RHIC energies in 2025. The eagerly-awaited sPHENIX quarkonium measurements will increase, for example, the precision of the bottomonium results, resolving the three $\YS$ states in a kinematic range that will be comparable to the LHC one. 

At LHC energies, ALICE 3~\cite{ALICE:2022wwr}, a new proposed experiment at CERN LHC after LS4 (LHC long shutdown 4 currently foreseen in 2034-2035), aims to improve even further the quality of quarkonium results, thanks to excellent vertexing and particle identification capabilities. The plan is to perform quarkonium spectroscopy studies adressing not only the P-wave quarkonium states but also the $\eta_c$ and $\eta_b$, so far never measured. 

While experiments at RHIC and LHC explore the QCD phase diagram at baryochemical potential ($\mu_{B}$) close to zero, a new line of investigation will be opened by the NA60+~\cite{NA60:2022sze}, a fixed-target experiment proposed at CERN SPS. NA60+ aims to study charmonium in the so far unexplored high-$\mu_{B}$ region ($\mu_{B}\sim$ 220 - 550 MeV), via a Pb beam energy scan between 6 $< \sqrtsNN<$ 17.3 GeV. The large luminosity exploited will allow a very precise measurement of the $\jpsi$ modification at all energies and by measuring the temperature of the system, via thermal dimuons, it will be possible to correlate it to the onset of the charmonium melting. 

The high-$\mu_{B}$ region of the QCD phase diagram will be covered also by the CBM experiment\,\cite{CBM:2016kpk} at FAIR. The collision energies ($\sqrtsNN<5$\, GeV) will allow the study of sub-threshold quarkonium production in Au-Au collisions.
The theoretical concepts discussed here for the collider energies (quarkonium melting in QGP, charm-quark thermalization) will not apply for such low energies but transport model calculations are available \,\cite{Cassing:2000vx}.

An exciting future is ahead, with quarkonium physics still having a central role in the experimental programs of the upgrades of existing experiments or of the future ones. New very high precision results will enrich the existing measurements and, in parallel, advancements in theoretical models will contribute to a more comprehensive understanding of the strong interactions and the fate of bound quark-antiquark systems.

% Summary Points
\begin{summary}[SUMMARY POINTS]
\begin{enumerate}
\item Quarkonium is a unique signature of QGP formation. The $\jpsi$ has been studied since the beginning of the heavy-ion program, and its measurements now span a broad kinematic and collision energy range. 

\item The $\jpsi$ production in Pb--Pb collisions exhibits significant modifications relative to pp.
In particular, the LHC $\raa$ results show a strong $\pT$ dependence, being suppressed at high $\pT$ but rising at low $\pT$, even exceeding unity at midrapidity. This behaviour was not present at lower energies and can be explained assuming suppression and (re)generation mechanisms at play.

\item The loosely bound $\psip$ exhibits similar features to the $\jpsi$, but with an $\raa$ reduction roughly twice as strong, indicating a hierarchy in nuclear modifications.

\item A clear sequential suppression is observed for bottomonia, with the reduction becoming stronger from the $\YiS$ to the $\YiiiS$, pointing to a dominant role of suppression mechanisms, with limited contribution from (re)generation processes.

\item The role of $J/\psi$ (re)generation at LHC is confirmed by the observation of a significant $\vtwo$, due to the collective flow inherited from thermalized charm quarks. In contrast, the $\YiS$ $\vtwo$ is consistent with zero, although with large uncertainties. 

\item Theoretical models, such as the statistical hadronization model and those based on transport equations successfully describe many aspects of quarkonium behavior. 

\item The precision achieved in quarkonium measurements is already very good, however, several open questions remain. Addressing these questions will require future data-taking at existing experiments, some of which will undergo significant upgrades soon, and measurements at the next generation of experiments such as ALICE 3 and NA60+, covering the QCD phase diagram from low to high $\mu_B$.
\end{enumerate}
\end{summary}

\section*{DISCLOSURE STATEMENT}
The authors are not aware of any affiliations, memberships, funding, or financial holdings that might be perceived as affecting the objectivity of this review. 

\section*{ACKNOWLEDGMENTS}
We acknowledge discussions with N. Brambilla, P.-B. Gossiaux, T. Magorsch, P. Petreczky, and R. Rapp.

% References

%\bibliographystyle{ar-style5} %w. article style
%\bibliography{quarkonium}

\bibliography{quarkonium.bbl}

\begin{thebibliography}{152}
\expandafter\ifx\csname natexlab\endcsname\relax\def\natexlab#1{#1}\fi

\bibitem{Matsui:1986dk}
Matsui T, Satz H.
\newblock \textit{Phys. Lett. B} 178:416--422 (1986)

\bibitem{Busza:2018rrf}
Busza W, Rajagopal K, van~der Schee W.
\newblock \textit{Ann. Rev. Nucl. Part. Sci.} 68:339--376 (2018)

\bibitem{Harris:2024aov}
Harris JW, M\"uller B.
\newblock \textit{Eur. Phys. J. C} 84(3):247 (2024)

\bibitem{Digal:2001ue}
Digal S, Petreczky P, Satz H.
\newblock \textit{Phys. Rev. D} 64:094015 (2001)

\bibitem{Karsch:2005nk}
Karsch F, Kharzeev D, Satz H.
\newblock \textit{Phys. Lett. B} 637:75--80 (2006)

\bibitem{NA50:2004sgj}
Alessandro B, et~al.
\newblock \textit{Eur. Phys. J. C} 39:335--345 (2005)

\bibitem{CERNpressrelease}
{CERN News}.
\newblock New state of matter created at cern,
  https://home.cern/news/press-release/cern/new-state-matter-created-cern
  (2000)

\bibitem{NA50:2006yzz}
Alessandro B, et~al.
\newblock \textit{Eur. Phys. J. C} 49:559--567 (2007)

\bibitem{BraunMunzinger:2000csl}
Braun-Munzinger P, Stachel J.
\newblock \textit{Phys. Lett. B} 490:196--202 (2000)

\bibitem{Thews:2000rj}
Thews RL, Schroedter M, Rafelski J.
\newblock \textit{Phys. Rev. C} 63:054905 (2001)

\bibitem{Andronic:2006ky}
Andronic A, Braun-Munzinger P, Redlich K, Stachel J.
\newblock \textit{Nucl. Phys. A} 789:334--356 (2007)

\bibitem{Lansberg:2019adr}
Lansberg JP.
\newblock \textit{Phys. Rept.} 889:1--106 (2020)

\bibitem{Boyd:2023ybk}
Boyd J, Thapa S, Strickland M.
\newblock \textit{Phys. Rev. D} 108(9):094024 (2023)

\bibitem{ALICE:2021edd}
Acharya S, et~al.
\newblock \textit{JHEP} 03:190 (2022)

\bibitem{LHCb:2021pyk}
Aaij R, et~al.
\newblock \textit{JHEP} 11:181 (2021)

\bibitem{Mocsy:2013syh}
Mocsy A, Petreczky P, Strickland M.
\newblock \textit{Int. J. Mod. Phys. A} 28:1340012 (2013)

\bibitem{He:2022ywp}
He M, van Hees H, Rapp R.
\newblock \textit{Prog. Part. Nucl. Phys.} 130:104020 (2023)

\bibitem{Bazavov:2023dci}
Bazavov A, Hoying D, Larsen RN, Mukherjee S, Petreczky P, et~al.
\newblock \textit{Phys. Rev. D} 109(7):074504 (2024)

\bibitem{Kluberg:2009wc}
Kluberg L, Satz H.
\newblock \textit{Landolt-Börnstein I} 23 (2010)

\bibitem{Rothkopf:2019ipj}
Rothkopf A.
\newblock \textit{Phys. Rept.} 858:1--117 (2020)

\bibitem{Du:2019tjf}
Du X, Liu SYF, Rapp R.
\newblock \textit{Phys. Lett. B} 796:20--25 (2019)

\bibitem{ALICE:2022wpn}
Acharya S, et~al.
\newblock \textit{Eur. Phys. J. C} 84(8):813 (2024)

\bibitem{Dong:2019byy}
Dong X, Lee YJ, Rapp R.
\newblock \textit{Ann. Rev. Nucl. Part. Sci.} 69:417--445 (2019)

\bibitem{Apolinario:2022vzg}
Apolin\'ario L, Lee YJ, Winn M.
\newblock \textit{Prog. Part. Nucl. Phys.} 127:103990 (2022)

\bibitem{Das:2024vac}
Das SK, Torres-Rincon JM, Rapp R.
\newblock \textit{Phys. Rept.} 1129-1131:1--53 (2025)

\bibitem{Zhao:2020jqu}
Zhao J, Zhou K, Chen S, Zhuang P.
\newblock \textit{Prog. Part. Nucl. Phys.} 114:103801 (2020)

\bibitem{Rapp:2008tf}
Rapp R, Blaschke D, Crochet P.
\newblock \textit{Prog. Part. Nucl. Phys.} 65:209--266 (2010)

\bibitem{Andronic:2024oxz}
Andronic A, et~al.
\newblock \textit{Eur. Phys. J. A} 60(4):88 (2024)

\bibitem{BraunMunzinger:2009dzl}
Braun-Munzinger P, Stachel J.
\newblock \textit{Landolt-B{\"o}rnstein I} 23:424 (2010)

\bibitem{Andronic:2021erx}
Andronic A, Braun-Munzinger P, K\"ohler MK, Mazeliauskas A, Redlich K, et~al.
\newblock \textit{JHEP} 07:035 (2021)

\bibitem{Andronic:2017pug}
Andronic A, Braun-Munzinger P, Redlich K, Stachel J.
\newblock \textit{Nature} 561(7723):321--330 (2018)

\bibitem{HotQCD:2018pds}
Bazavov A, et~al.
\newblock \textit{Phys. Lett. B} 795:15--21 (2019)

\bibitem{Borsanyi:2020fev}
Borsanyi S, Fodor Z, Guenther JN, Kara R, Katz SD, et~al.
\newblock \textit{Phys. Rev. Lett.} 125(5):052001 (2020)

\bibitem{Cleymans:1990mn}
Cleymans J, Redlich K, Suhonen E.
\newblock \textit{Z. Phys. C} 51:137--141 (1991)

\bibitem{Gorenstein:2000ck}
Gorenstein MI, Kostyuk A, Stoecker H, Greiner W.
\newblock \textit{Phys. Lett. B} 509:277--282 (2001)

\bibitem{Andronic:2019wva}
Andronic A, Braun-Munzinger P, K\"ohler MK, Redlich K, Stachel J.
\newblock \textit{Phys. Lett. B} 797:134836 (2019)

\bibitem{dEnterria:2020dwq}
d'Enterria D, Loizides C.
\newblock \textit{Ann. Rev. Nucl. Part. Sci.} 71:315--344 (2021)

\bibitem{Braun-Munzinger:2000uqj}
Braun-Munzinger P, Redlich K.
\newblock \textit{Eur. Phys. J. C} 16:519--525 (2000)

\bibitem{Andronic:2023tui}
Andronic A, Braun-Munzinger P, Brun\ss{}en H, Crkovsk\'a J, Stachel J, et~al.
\newblock \textit{JHEP} 10:229 (2024)

\bibitem{Andronic:2022ucg}
Andronic A, Braun-Munzinger P, Redlich K, Stachel J.
\newblock \textit{Acta Phys. Polon. Supp.} 16(1):1--A107 (2023)

\bibitem{Grandchamp:2001pf}
Grandchamp L, Rapp R.
\newblock \textit{Phys. Lett. B} 523:60--66 (2001)

\bibitem{Wu:2024gil}
Wu B, Rapp R.
\newblock \textit{Universe} 10(6):244 (2024)

\bibitem{Tang:2024dkz}
Tang Z, Mukherjee S, Petreczky P, Rapp R.
\newblock {arXiv:2411.09132} (2024)

\bibitem{Larsen:2019zqv}
Larsen R, Meinel S, Mukherjee S, Petreczky P.
\newblock \textit{Phys. Lett. B} 800:135119 (2020)

\bibitem{Brambilla:2024tqg}
Brambilla N, Magorsch T, Strickland M, Vairo A, Vander~Griend P.
\newblock \textit{Phys. Rev. D} 109(11):114016 (2024)

\bibitem{Delorme:2024rdo}
Delorme S, Katz R, Gousset T, Gossiaux PB, Blaizot JP.
\newblock \textit{JHEP} 06:060 (2024)

\bibitem{Brambilla:2023hkw}
Brambilla N, Escobedo MA, Islam A, Strickland M, Tiwari A, et~al.
\newblock \textit{Phys. Rev. D} 108(1):L011502 (2023)

\bibitem{Rapp:2017chc}
Rapp R, Du X.
\newblock \textit{Nucl. Phys. A} 967:216--224 (2017)

\bibitem{Akamatsu:2020ypb}
Akamatsu Y.
\newblock \textit{Prog. Part. Nucl. Phys.} 123:103932 (2022)

\bibitem{Brambilla:1999xf}
Brambilla N, Pineda A, Soto J, Vairo A.
\newblock \textit{Nucl. Phys. B} 566:275 (2000)

\bibitem{ALICE:2023gco}
Acharya S, et~al.
\newblock \textit{Phys. Lett. B} 849:138451 (2024)

\bibitem{ALICE:2021rxa}
Acharya S, et~al.
\newblock \textit{JHEP} 01:174 (2022)

\bibitem{Zhao:2011cv}
Zhao X, Rapp R.
\newblock \textit{Nucl. Phys. A} 859:114--125 (2011)

\bibitem{Zhou:2014kka}
Zhou K, Xu N, Xu Z, Zhuang P.
\newblock \textit{Phys. Rev. C} 89(5):054911 (2014)

\bibitem{Klasen:2023uqj}
Klasen M, Paukkunen H.
\newblock \textit{Ann. Rev. Nucl. Part. Sci.} 74:49--87 (2024)

\bibitem{He:2019tik}
He M, Rapp R.
\newblock \textit{Phys. Lett. B} 795:117--121 (2019)

\bibitem{ALICE:2023sgl}
Acharya S, et~al.
\newblock \textit{JHEP} 12:086 (2023)

\bibitem{Ferreiro:2012rq}
Ferreiro EG.
\newblock \textit{Phys. Lett. B} 731:57--63 (2014)

\bibitem{Ferreiro:2018wbd}
Ferreiro EG, Lansberg JP.
\newblock \textit{JHEP} 10:094 (2018), [Erratum: JHEP 03, 063 (2019)]

\bibitem{Song:2023zma}
Song T, Aichelin J, Zhao J, Gossiaux PB, Bratkovskaya E.
\newblock \textit{Phys. Rev. C} 108(5):054908 (2023)

\bibitem{Wolschin:2020kwt}
Wolschin G.
\newblock \textit{Int. J. Mod. Phys. A} 35(29):2030016 (2020)

\bibitem{Blaizot:2021xqa}
Blaizot JP, Escobedo MA.
\newblock \textit{Phys. Rev. D} 104(5):054034 (2021)

\bibitem{Yao:2020xzw}
Yao X, Ke W, Xu Y, Bass SA, M\"uller B.
\newblock \textit{JHEP} 01:046 (2021)

\bibitem{Villar:2022sbv}
Villar DYA, Zhao J, Aichelin J, Gossiaux PB.
\newblock \textit{Phys. Rev. C} 107(5):054913 (2023)

\bibitem{Chen:2024iil}
Chen G, Chen B, Zhao J.
\newblock \textit{Eur. Phys. J. C} 84(8):869 (2024)

\bibitem{CMS:2023lfu}
Tumasyan A, et~al.
\newblock \textit{Phys. Rev. Lett.} 133(2):022302 (2024)

\bibitem{Faccioli:2010kd}
Faccioli P, Lourenco C, Seixas J, Wohri HK.
\newblock \textit{Eur. Phys. J. C} 69:657--673 (2010)

\bibitem{Andronic:2015wma}
Andronic A, et~al.
\newblock \textit{Eur. Phys. J. C} 76(3):107 (2016)

\bibitem{NA50:2006rdp}
Alessandro B, et~al.
\newblock \textit{Eur. Phys. J. C} 48:329 (2006)

\bibitem{NA50:2003fvu}
Alessandro B, et~al.
\newblock \textit{Eur. Phys. J. C} 33:31--40 (2004)

\bibitem{NA60:2010wey}
Arnaldi R, et~al.
\newblock \textit{Phys. Lett. B} 706:263--267 (2012)

\bibitem{Eskola:2016oht}
Eskola KJ, Paakkinen P, Paukkunen H, Salgado CA.
\newblock \textit{Eur. Phys. J. C} 77(3):163 (2017)

\bibitem{Kovarik:2015cma}
Kovarik K, et~al.
\newblock \textit{Phys. Rev. D} 93(8):085037 (2016)

\bibitem{Iancu:2003xm}
Iancu E, Venugopalan R.
\newblock {hep-ph/0303204} (2003)

\bibitem{Arleo:2012rs}
Arleo F, Peigne S.
\newblock \textit{JHEP} 03:122 (2013)

\bibitem{NA50:2003pvd}
Alessandro B, et~al.
\newblock \textit{Phys. Lett. B} 553:167--178 (2003)

\bibitem{PHENIX:2012czk}
Adare A, et~al.
\newblock \textit{Phys. Rev. C} 87(3):034904 (2013)

\bibitem{Blaizot:1987ha}
Blaizot JP, Ollitrault JY.
\newblock \textit{Phys. Lett. B} 199:499--503 (1987)

\bibitem{Hufner:1989fn}
Hufner J, Povh B, Gardner S.
\newblock \textit{Phys. Lett. B} 238:103--107 (1990)

\bibitem{Kharzeev:1999bh}
Kharzeev D, Thews RL.
\newblock \textit{Phys. Rev. C} 60:041901 (1999)

\bibitem{ALICE:2019qie}
Acharya S, et~al.
\newblock \textit{Phys. Lett. B} 806:135486 (2020)

\bibitem{ATLAS:2017prf}
Aaboud M, et~al.
\newblock \textit{Eur. Phys. J. C} 78(3):171 (2018)

\bibitem{CMS:2022wfi}
Tumasyan A, et~al.
\newblock \textit{Phys. Lett. B} 835:137397 (2022)

\bibitem{LHCb:2018psc}
Aaij R, et~al.
\newblock \textit{JHEP} 11:194 (2018), [Erratum: JHEP 02, 093 (2020)]

\bibitem{PHENIX:2022nrm}
Acharya UA, et~al.
\newblock \textit{Phys. Rev. C} 105(6):064912 (2022)

\bibitem{PHENIX:2019brm}
Acharya U, et~al.
\newblock \textit{Phys. Rev. C} 102(1):014902 (2020)

\bibitem{STAR:2021zvb}
Abdallah M, et~al.
\newblock \textit{Phys. Lett. B} 825:136865 (2022)

\bibitem{Albacete:2017qng}
Albacete JL, et~al.
\newblock \textit{Nucl. Phys. A} 972:18--85 (2018)

\bibitem{Vogt:2019jxd}
Vogt R.
\newblock {hep-ph:1908.11534} (2019)

\bibitem{Strickland:2024oat}
Strickland M, Thapa S, Vogt R.
\newblock \textit{Phys. Rev. D} 109(9):096016 (2024)

\bibitem{Du:2018wsj}
Du X, Rapp R.
\newblock \textit{JHEP} 03:015 (2019)

\bibitem{ALICE:2016sdt}
Adam J, et~al.
\newblock \textit{JHEP} 06:050 (2016)

\bibitem{ALICE:2014cgk}
Abelev BB, et~al.
\newblock \textit{JHEP} 12:073 (2014)

\bibitem{ALICE:2020vjy}
Acharya S, et~al.
\newblock \textit{JHEP} 07:237 (2020)

\bibitem{Arleo:1999af}
Arleo F, Gossiaux PB, Gousset T, Aichelin J.
\newblock \textit{Phys. Rev. C} 61:054906 (2000)

\bibitem{McGlinchey:2012bp}
McGlinchey DC, Frawley AD, Vogt R.
\newblock \textit{Phys. Rev. C} 87(5):054910 (2013)

\bibitem{LHCb:2016vqr}
Aaij R, et~al.
\newblock \textit{JHEP} 03:133 (2016)

\bibitem{PHENIX:2016vmz}
Adare A, et~al.
\newblock \textit{Phys. Rev. C} 95(3):034904 (2017)

\bibitem{Ma:2017rsu}
Ma YQ, Venugopalan R, Watanabe K, Zhang HF.
\newblock \textit{Phys. Rev. C} 97(1):014909 (2018)

\bibitem{Ferreiro:2014bia}
Ferreiro EG.
\newblock \textit{Phys. Lett. B} 749:98--103 (2015)

\bibitem{Arnaldi:2008zz}
Arnaldi R.
\newblock \textit{J. Phys. G} 35:104133 (2008)

\bibitem{Kharzeev:1996yx}
Kharzeev D, Lourenco C, Nardi M, Satz H.
\newblock \textit{Z. Phys. C} 74:307--318 (1997)

\bibitem{Capella:2000zp}
Capella A, Ferreiro EG, Kaidalov AB.
\newblock \textit{Phys. Rev. Lett.} 85:2080--2083 (2000)

\bibitem{Blaizot:1996nq}
Blaizot JP, Ollitrault JY.
\newblock \textit{Phys. Rev. Lett.} 77:1703--1706 (1996)

\bibitem{PHENIX:2006gsi}
Adare A, et~al.
\newblock \textit{Phys. Rev. Lett.} 98:232301 (2007)

\bibitem{PHENIX:2011img}
Adare A, et~al.
\newblock \textit{Phys. Rev. C} 84:054912 (2011)

\bibitem{STAR:2013eve}
Adamczyk L, et~al.
\newblock \textit{Phys. Rev. C} 90(2):024906 (2014)

\bibitem{STAR:2019fge}
Adam J, et~al.
\newblock \textit{Phys. Lett. B} 797:134917 (2019)

\bibitem{CMS:2017uuv}
Sirunyan AM, et~al.
\newblock \textit{Eur. Phys. J. C} 78(6):509 (2018), [Erratum: Eur.Phys.J.C 83,
  145 (2023)]

\bibitem{ATLAS:2018xms}
Aaboud M, et~al.
\newblock \textit{Eur. Phys. J. C} 78(9):784 (2018)

\bibitem{ALICE:2023hou}
Acharya S, et~al.
\newblock \textit{JHEP} 02:066 (2024)

\bibitem{ALICE:2021gpt}
Acharya S, et~al.
\newblock \textit{Eur. Phys. J. C} 81(8):712 (2021)

\bibitem{ALICE:2016flj}
Adam J, et~al.
\newblock \textit{Phys. Lett. B} 766:212--224 (2017)

\bibitem{He:2021zej}
He M, Wu B, Rapp R.
\newblock \textit{Phys. Rev. Lett.} 128(16):162301 (2022)

\bibitem{Capellino:2023cxe}
Capellino F, Dubla A, Floerchinger S, Grossi E, Kirchner A, Masciocchi S.
\newblock \textit{Phys. Rev. D} 108(11):116011 (2023)

\bibitem{ALICE:2017quq}
Acharya S, et~al.
\newblock \textit{Phys. Rev. Lett.} 119(24):242301 (2017)

\bibitem{CMS:2023mtk}
Tumasyan A, et~al.
\newblock \textit{JHEP} 10:115 (2023)

\bibitem{PHENIX:2024axj}
Abdulameer NJ, et~al.
\newblock {arXiv:2409.12756} (2024)

\bibitem{STAR:2012jzy}
Adamczyk L, et~al.
\newblock \textit{Phys. Rev. Lett.} 111(5):052301 (2013)

\bibitem{Noronha-Hostler:2016eow}
Noronha-Hostler J, Betz B, Noronha J, Gyulassy M.
\newblock \textit{Phys. Rev. Lett.} 116(25):252301 (2016)

\bibitem{ALICE:2020iev}
Acharya S, et~al.
\newblock \textit{Phys. Lett. B} 815:136146 (2021)

\bibitem{ALICE:2022dyy}
Acharya S, et~al.
\newblock \textit{Phys. Rev. Lett.} 131(4):042303 (2023)

\bibitem{ALICE:2019aid}
Acharya S, et~al.
\newblock \textit{Phys. Rev. Lett.} 125(1):012301 (2020)

\bibitem{LHCb:2023apa}
Aaij R, et~al.
\newblock \textit{Phys. Rev. Lett.} 132(10):102302 (2024)

\bibitem{ALICE:2022jeh}
Acharya S, et~al.
\newblock \textit{Phys. Rev. Lett.} 132(4):042301 (2024)

\bibitem{Du:2015wha}
Du X, Rapp R.
\newblock \textit{Nucl. Phys. A} 943:147--158 (2015)

\bibitem{CMS:2012gvv}
Chatrchyan S, et~al.
\newblock \textit{Phys. Rev. Lett.} 109:222301 (2012), [Erratum: Phys.Rev.Lett.
  120, 199903 (2018)]

\bibitem{CMS:2017ycw}
Sirunyan AM, et~al.
\newblock \textit{Phys. Rev. Lett.} 120(14):142301 (2018)

\bibitem{CMS:2018zza}
Sirunyan AM, et~al.
\newblock \textit{Phys. Lett. B} 790:270--293 (2019)

\bibitem{ATLAS:2022exb}
Aad G, et~al.
\newblock \textit{Phys. Rev. C} 107(5):054912 (2023)

\bibitem{ALICE:2020wwx}
Acharya S, et~al.
\newblock \textit{Phys. Lett. B} 822:136579 (2021)

\bibitem{STAR:2022rpk}
Aboona B, et~al.
\newblock \textit{Phys. Rev. Lett.} 130(11):112301 (2023)

\bibitem{Du:2017qkv}
Du X, He M, Rapp R.
\newblock \textit{Phys. Rev. C} 96(5):054901 (2017)

\bibitem{CMS:2013jsu}
Chatrchyan S, et~al.
\newblock \textit{JHEP} 04:103 (2014)

\bibitem{Acharya:2019hlv}
Acharya S, et~al.
\newblock \textit{Phys. Rev. Lett.} 123(19):192301 (2019)

\bibitem{Sirunyan:2020qec}
Sirunyan AM, et~al.
\newblock \textit{Phys. Lett. B} 819:136385 (2021)

\bibitem{ALICE:2019lga}
Acharya S, et~al.
\newblock \textit{JHEP} 02:041 (2020)

\bibitem{ALICE:2018mml}
Acharya S, et~al.
\newblock \textit{JHEP} 07:160 (2018)

\bibitem{ALICE:2013snh}
Abelev BB, et~al.
\newblock \textit{JHEP} 02:073 (2014)

\bibitem{ALICE:2015sru}
Adam J, et~al.
\newblock \textit{JHEP} 06:055 (2015)

\bibitem{Belle:2003nnu}
Choi SK, et~al.
\newblock \textit{Phys. Rev. Lett.} 91:262001 (2003)

\bibitem{CDF:2006ocq}
Abulencia A, et~al.
\newblock \textit{Phys. Rev. Lett.} 98:132002 (2007)

\bibitem{LHCb:2013kgk}
Aaij R, et~al.
\newblock \textit{Phys. Rev. Lett.} 110:222001 (2013)

\bibitem{Tornqvist:2004qy}
Tornqvist NA.
\newblock \textit{Phys. Lett. B} 590:209--215 (2004)

\bibitem{Maiani:2004vq}
Maiani L, Piccinini F, Polosa AD, Riquer V.
\newblock \textit{Phys. Rev. D} 71:014028 (2005)

\bibitem{Hanhart:2011jz}
Hanhart C, Kalashnikova YS, Nefediev AV.
\newblock \textit{Eur. Phys. J. A} 47:101--110 (2011)

\bibitem{LHCb:2024bpb}
Aaij R, et~al.
\newblock \textit{Phys. Rev. Lett.} 132(24):242301 (2024)

\bibitem{CMS:2021znk}
Sirunyan AM, et~al.
\newblock \textit{Phys. Rev. Lett.} 128(3):032001 (2022)

\bibitem{ALICE:2022wwr}
ALICE.
\newblock {Collaboration, Letter of intent for ALICE 3: A next-generation
  heavy-ion experiment at the LHC, arXiv:2211.02491, CERN-LHCC-2022-009,
  LHCC-I-038} (2022)

\bibitem{NA60:2022sze}
Ahdida C, et~al.
\newblock {Letter of Intent: the NA60+ experiment, arXiv:2212.14452,
  CERN-SPSC-2022-036 / SPSC-I-259} (2022); CERN-SPSC-2025-023 / SPSC-P-373 (2025)

\bibitem{CBM:2016kpk}
Ablyazimov T, et~al.
\newblock \textit{Eur. Phys. J. A} 53(3):60 (2017)

\bibitem{Cassing:2000vx}
Cassing W, Bratkovskaya EL, Sibirtsev A.
\newblock \textit{Nucl. Phys. A} 691:753--778 (2001)

\end{thebibliography}

\end{document}